\documentclass[11pt,prd,onecolumn,amsmath,amssymb,aps,floats,floatfix,nofootinbib, superscriptaddress]{revtex4-2}
\usepackage[colorlinks=true,urlcolor=blue,anchorcolor=blue,citecolor=blue,filecolor=blue,linkcolor=blue,menucolor=blue,linktocpage=true]{hyperref} % should be commented out if the tex file will be compiled with latex in arXiv!!! (pdflatex is fine)

\usepackage[inline]{enumitem}
\usepackage[multidot]{grffile}  % allow the name of figures to include dots
\usepackage{dcolumn}
\usepackage{bm}
\usepackage{amsmath}
\usepackage{amsfonts}
\usepackage{amssymb}
\usepackage{color}
\usepackage[table]{xcolor}
\usepackage{float}
\usepackage{latexsym}
\usepackage{slashed} % slash mark
\usepackage{pstricks}
\usepackage{indentfirst}
\usepackage{mathrsfs}
\usepackage{multirow}
\usepackage{epsfig,psfrag}
\usepackage{graphicx}
\usepackage{enumitem}
\usepackage{geometry,amssymb,yfonts}
\usepackage{yhmath}
\usepackage{anysize}
\usepackage{subfigure}
\usepackage{mathtools}
\usepackage{setspace} % spacing
\usepackage[utf8]{inputenc} % accept utf-8 input encoding
\usepackage[scientific-notation=true]{siunitx} % comprehensive units

\usepackage{url}
\usepackage{makecell}

\graphicspath{{fig/}}

\allowdisplaybreaks % allow eqnarray breaks

\definecolor{myred}{rgb}{.75,0,0}

\usepackage{booktabs}
\usepackage{verbatim}

\begin{document}
\title{Collapsing domain walls with $\mathbb{Z}_2$-violating coupling to thermalized fermions and their impact on gravitational wave detections}
\author{Qing-Quan Zeng}
\affiliation{School of Physics, Sun Yat-Sen University, Guangzhou 510275, China}
\author{Xi He}
\affiliation{School of Physics, Sun Yat-Sen University, Guangzhou 510275, China}
\author{Zhao-Huan Yu}\email[Corresponding author. ]{yuzhaoh5@mail.sysu.edu.cn}
\affiliation{School of Physics, Sun Yat-Sen University, Guangzhou 510275, China}
\author{Jiaming Zheng}\email[Corresponding author. ]{zhengjm3@gmail.com}

\begin{abstract}
We study the dynamics of domain walls formed through the spontaneous breaking of an approximate $\mathbb{Z}_2$ symmetry in a scalar field, focusing on their collapse under the influence of quantum and thermal corrections induced by a $\mathbb{Z}_2$-violating Yukawa coupling to Dirac fermions in the thermal bath. The thermal effects make the potential bias between the true and false vacua dependent on the temperature and may lead to notable variations in the annihilation temperature of domain walls, in addition to the shift caused by temperature-independent quantum corrections. These modifications could substantially alter the gravitational wave spectrum produced by collapsing domain walls, potentially providing observable signatures for future gravitational wave detection experiments.
\end{abstract}

\maketitle
\tableofcontents

\clearpage

\section{Introduction}

Since the ground-breaking observations of gravitational wave\,(GW) events by the %Laser Interferometer Gravitational-Wave Observatory Scientific Collaboration and 
Laser Interferometer Gravitational Wave Observatory (LIGO)-Virgo Collaboration\,\cite{LIGOScientific:2016aoc,LIGOScientific:2017vwq}, the possibility of detecting gravitational wave signals in the early Universe has drawn tremendous attention in physics research. 
Various well-motivated hypothesized processes in the primordial Universe, such as cosmic inflation, cosmological phase transitions, and the evolution of topological defects, naturally source stochastic gravitational wave backgrounds\,(SGWBs) that may be accessible by future GW experiments\,\cite{Caprini:2018mtu, Renzini:2022alw}, e.g., ground-based interferometers in the frequency band $10\text{--}10^3~\si{Hz}$\,\cite{KAGRA:2013rdx,LIGOScientific:2016wof}, pulsar timing arrays (PTAs) in $10^{-9}\text{--}10^{-7}~\si{Hz}$\,\cite{NANOGRAV:2018hou,Lentati:2015qwp,Shannon:2015ect,Hobbs:2009yy,Janssen:2014dka}, and future space-borne interferometers in $10^{-4}\text{--}10^{-1}~\si{Hz}$\,\cite{LISA:2017pwj,Ruan:2018tsw,Liang:2021bde}.
%In particular,  There are two main sources of the stochastic gravitational waves \cite{Renzini:2022alw}, one is celestial movements, and the other one is the primary cosmos background, such as cosmic inflation, non-linear phenomena, primordial black holes, and so on.
%Naturally, the stochastic gravitational waves background is attractive for scientists to conduct various types of gravitational waves experiments to detect it in different frequency bands.
Intriguingly, several PTA collaborations have reported positive evidence recently for a nanohertz SGWB
\cite{NANOGrav:2023gor, EPTA:2023fyk, Reardon:2023gzh, Xu:2023wog} that may be interpreted as the smoking gun of new physics beyond the standard model%
\,\cite{NANOGrav:2023hvm, EPTA:2023xxk}.
%Moreover, a new technical method, resonant electromagnetic gravitational-wave detectors \cite{Herman:2020wao, Herman:2022fau}, has also been raised to detect gravitational waves at higher frequency band.

%%%%%%%%%%%%%%%%%%%%%%%%%%%%%%%%%%
In various extensions of the standard model (SM), such as the grand unification theories, the high energy theory has a large symmetry group that must be broken by heavy Higgs fields in the early Universe. Depending on the symmetry-breaking pattern, the corresponding phase transition may produce topological defects such as domain walls\,(DWs), cosmic strings, and monopoles through the Kibble mechanism\,\cite{Kibble:1976sj}.  
% Topological defects are also one of sources from the phase transitions of fields, including cosmic strings \cite{Battye:1997ji}, domain walls \cite{Hiramatsu:2013qaa, Bian:2022qbh}, monopoles \cite{Martin:1996cp} and textures. Therefore, it is significant to research the gravitational waves emitted from domain walls.
In this work, we shall focus on the evolution of domain walls, the two-dimensional topological defects formed in the scalar field that spontaneously breaks a discrete symmetry\,\cite{Kibble:1976sj, Vilenkin:1981zs, Vachaspati:2000cq, Vilenkin:2000jqa}. 
%While the symmetric group of a field is spontaneously broken into the other group, domain walls will occur if their quotient group has two or more disconnected components \cite{Kibble:1976sj, Vilenkin:1981zs, Vachaspati:2000cq, Vilenkin:2000jqa}.

Stable DWs are cosmologically problematic because their energy density falls slower than those of matter and radiation and would soon overclose the Universe\,\cite{Zeldovich:1974uw, Press:1989yh}. 
However, global symmetries are generally not expected to be exact since they are violated explicitly at least by quantum gravity effects\,\cite{Abbott:1989jw,Coleman:1989zu}.
If a global discrete symmetry is only weakly broken, DWs are still produced during the phase transition. The symmetry-violating effect, usually represented by a bias term in the scalar potential%
\,\cite{Hiramatsu:2010yz,Kawasaki:2011vv,Gelmini:1988sf,Kadota:2015dza,Borah:2022wdy}, %
%\footnote{It may come from higher dimensional gauge invariant but parity-breaking operators, Planck-scale-suppressed nonrenormalizable operators \cite{Borah:2022wdy}, 
%odd-power terms \cite{Hiramatsu:2010yz, Kawasaki:2011vv}, temperature corrections \cite{Gelmini:1988sf}, the coupling with singlet superfields \cite{Kadota:2015dza}, 
%or other sources.
%}
renders the DWs unstable\,\cite{Gelmini:1988sf, Coulson:1995nv, Hiramatsu:2010yz} by sourcing pressure that tends to reduce the domains of the false vacuum.
The collapsing DWs may generate a significant amount of GWs that further accelerates this process\,\cite{Vilenkin:1981zs} and forms an SGWB that may be probed by GW experiments\,\cite{Gleiser:1998na,Hiramatsu:2010yz,Hiramatsu:2013qaa,Nakayama:2016gxi,Saikawa:2017hiv,Gelmini:2020bqg,Zhang:2023nrs}.

In passing, we note that there exist alternative solutions to the cosmological DW problem. For instance, the DWs may be destroyed by primordial black holes\,\cite{Stojkovic:2005zh}, by cosmic string loops nucleated on the wall\,\cite{Dunsky:2021tih}, or by cosmic strings that bound them\,\cite{Lazarides:1982tw}, depending on the sequence symmetry breaking. However, this work is specifically concerned with DW collapse driven by discrete symmetry-violating terms.

The evolution of the DWs with a biased scalar potential has been studied thoroughly in the literature (see, e.g., Refs.~\cite{Nakayama:2016gxi,Saikawa:2017hiv} for reviews). However, the symmetry-breaking effect that generates the bias term in the scalar potential can also manifest in other interactions. For instance, the coupling between the scalar field and matter particles may also violate the discrete symmetry that stabilizes the DWs.
If the coupled particles are rare in the Universe, they only contribute to the DWs through radiative corrections\,\cite{Zhang:2023nrs} captured by the Coleman-Weinberg effective potential\,\cite{Coleman:1973jx} at the one-loop level.
On the other hand, if the particles are thermally populated, the asymmetric interaction between the scalar field and the particles sources energy gaps between different vacua that may cause the collapse of DWs. In this work, we estimate these effects with both the Coleman-Weinberg effective potential and the thermal effective potential and determine their relative importance compared to a bias term introduced in the bare scalar potential. Then, we compute the modified GW spectrum generated by the collapse of unstable DWs and investigate its detection perspective. 

The organization of this work is as follows.
In Sec.~\ref{sec:model}, we introduce the baseline model of this work that contains a real scalar field coupled to a Dirac fermionic field. Then we provide the corrections to the Lagrangian given by the Coleman-Weinberg effective potential\,\cite{Coleman:1973jx, Delaunay:2007wb} and the finite-temperature effective potential\,\cite{Quiros:1999jp}. 
%In our model, the potential bias between the local minima is variable. The formation of domain walls and the phase transition of real scalar field are described by the finite temperature field theory \cite{Quiros:1999jp}, and the bias will be changeable depended on temperature. The Coleman-Weinberg effective potentials \cite{Coleman:1973jx, Delaunay:2007wb, Quiros:1999jp} we considered are invariant as temperature. Therefore, we are interested to see how the fermions and the temperature corrections affect the annihilation of domain walls and the spectrum of gravitational waves in our study.
In Sec.~\ref{sec:evolution}, we solve the configuration equation of a DW and calculate its tension. Then, we analyze the evolution of the DW network driven by the bias term to obtain its annihilation temperature. The friction exerted by the fermions in the thermal bath on the DWs is also considered.
In Sec.~\ref{spectrum}, we demonstrate the GW spectra induced by the evolution of DWs with various benchmark parameters with and without the fermionic field and compare them to sensitivity curves of several GW detection experiments\,\cite{Lentati:2015qwp, NANOGRAV:2018hou, Shannon:2015ect, Hobbs:2009yy, Janssen:2014dka, KAGRA:2013rdx, LIGOScientific:2016wof, LISA:2017pwj, Liang:2021bde, Ruan:2018tsw}. 
In Sec.~\ref{sec:muR}, we examine the dependence of the DW annihilation temperature and the GW spectrum on the renormalization scale.
Sec.~\ref{sec:summary} is the summary.

\section{A Toy Model of Asymmetric Yukawa Coupling and Its Thermal Corrections}\label{sec:model}

We base our study on a toy model with a real scalar field $\phi$ that develops a domain wall configuration in the Universe and a Dirac fermionic field $f$ that couples to the scalar. The Lagrangian is
\begin{equation}
\mathcal{L}=\frac{1}{2}\partial_{\mu} \phi\partial^{\mu} \phi
+\mathrm{i}\bar{f}\gamma^\mu\partial_\mu f
- M_f( \phi)\bar{f}f-V_0( \phi),\end{equation}
where we have defined
\begin{equation}
M_f( \phi)\equiv m_f+y\phi,
\end{equation}
with a mass parameter $m_f$ and a Yukawa coupling $y$.
The scalar potential is taken as
\begin{equation}\label{eq:V0}
V_0(\phi) = \mu_1^3\phi -\frac{1}{2}\mu_\phi^2\phi^2+\frac{1}{3}\mu_3\phi^3+\frac{1}{4}\lambda_\phi\phi^4\, ,
\end{equation}
with $ \mu_\phi^2>0$ and $\lambda_\phi>0 $.

The $y$, $\mu_1^3$, and $ \mu_3 $ terms violate the $\mathbb{Z}_2$ symmetry of the Lagrangian, $\phi\rightarrow -\phi$, explicitly.
For simplicity, we set $\mu_1^3 = 0$, as this term can be eliminated by a redefinition of $\phi$.
When $y$ and $\mu_3$ are small, the approximate $\mathbb{Z}_2$ symmetry of the theory is spontaneously broken by the vacuum expectation value (VEV) of $ \phi $ and the corresponding second-order phase transition generates DWs through the Kibble mechanism\,\cite{Kibble:1976sj}.  

At the one-loop level, the radiative correction on the scalar potential is captured by the Coleman-Weinberg (CW) effective potentials $ V_{\mathrm{CW}}( \phi) $ \cite{Coleman:1973jx, Delaunay:2007wb} in the $ \overline{\mathrm{MS}} $ renormalization scheme, 
\begin{equation}
V_{\mathrm{CW}}( \phi)=\frac{m_{\phi}^4( \phi)}{64\pi^2}\,\mathrm{Re}\left[ \ln\frac{m_{\phi}^2( \phi)}{\mu_{\mathrm{R}}^2}-\frac{3}{2}\right]-\frac{M_f^4( \phi)}{16\pi^2}\left[ \ln\frac{M_f^2( \phi)}{\mu_{\mathrm{R}}^2}-\frac{3}{2}\right],
\label{eq:V_CW}
\end{equation}
where the field-dependent scalar mass squared is
\begin{equation}
m_{\phi}^2( \phi)\equiv\frac{\partial^2 V_0( \phi)}{\partial\phi^2}.
\end{equation}
We recall that for theoretical consistency, the parameters in the Lagrangian are also set at the renormalization scale $\mu_\mathrm{R}$, including the condition $\mu_1^3(\mu_\mathrm{R})=0$. When evaluating at another scale $\mu'_\mathrm{R}$, the scale dependence of the running parameters
%\footnote{\red{A non-zero $\mu_1^3$ may also be regenerated through renormalization group running at another scale $\mu'_R\neq \mu_R$.}}
and $V_{\mathrm{CW}}$ will cancel in such a way that physical quantities, such as the height of the minima of the potential remain unchanged within the order of perturbation theory\,\cite{Coleman:1973jx,Andreassen:2014eha}. We leave the detailed discussion of this point and its numerical robustness to Sec.~\ref{sec:muR}.
We fix the renormalization scale at $ \mu_{\mathrm{R}} = v_\phi$, where the VEV $v_\phi$ is evaluated by
\begin{equation}\label{eq:eq-6}
\left. \frac{\partial}{\partial\phi}\left[V_0( \phi)+ V_{\mathrm{CW}}( \phi) \right] \right|_{\phi=v_\phi}=0.
\end{equation}

We assume that the $f$ fermions and the $\phi$ scalar bosons are thermally produced in the early Universe, which is the case if they interact efficiently with SM particles, and the finite-temperature corrections\,\cite{Quiros:1999jp} to the effective scalar potential are given by
\begin{equation}\begin{split}
V_{\mathrm{T}}( \phi,T) =\,&
\frac{T^4}{2\pi^2}\,\mathrm{Re}\int_0^{+\infty}x^2\ln\left[ 1 - \mathrm{e}^{-\sqrt{x^2+m_{\phi}^2( \phi)/T^2}}\,\right] \mathrm{d}x
\\ &-\frac{2T^4}{\pi^2}\int_0^{+\infty}x^2\ln\left[ 1 + \mathrm{e}^{-\sqrt{x^2+M_f^2( \phi)/T^2}}\,\right] \mathrm{d}x\,.
\end{split}
\label{eq:V_T}
\end{equation}
The $\mathbb{Z}_2$-violating Yukawa coupling $y$ enters both the Coleman-Weinberg potential~\eqref{eq:V_CW} and the thermal effective potential~\eqref{eq:V_T} through the field-dependent mass $M_f(\phi)$ as an additional source for the potential bias.
Thus, the fully corrected effective potential is
\begin{equation}
V( \phi,T) = V_0( \phi)+ V_{\mathrm{CW}}( \phi)+ V_{\mathrm{T}}( \phi,T).
\end{equation}

In this work, we choose the following five parameters, 
$\mu_3$, $\lambda_\phi$, $y$, $m_f$, and $v_\phi$,
as free parameters. We restrict ourselves to small $\mathbb{Z}_2$-violating couplings, $\mu_3\ll v_\phi$ and $y\ll 1$, so that they only control the late time evolution of DWs while having little impact on the phase transition and the DW formation.
To illustrate our numerical results, we choose three benchmark points\,(BPs) of model parameters with nonzero $m_f$ and $y$, and denote them as ``BP$n$" ($ n=1,2,3$), which are listed in Table~\ref{tab:parameter}.
These parameters are specifically chosen for GW signals at various frequency bands that might be probed in future GW experiments.
For comparison, we also consider the evolution of DWs without introducing the fermion $f$, and the corresponding parameter sets with vanishing $m_f$ and $y$ are denoted as ``BP$n$ w/o $f$" in the following analysis.

\begin{table}[!t]
\caption{A list of benchmark points of the model.}\label{tab:parameter}
\belowrulesep=0pt
\aboverulesep=0pt
\centering
\renewcommand{\arraystretch}{1.5} 
\begin{tabular}{@{\hspace{1.3em}}c@{\hspace{3em}}c@{\hspace{3em}}c@{\hspace{3em}}c@{\hspace{3em}}c@{\hspace{3em}}c@{\hspace{1.3em}}}
\toprule[1.2pt]
&$v_\phi\,(\mathrm{GeV})$&$ \mu_3/v_\phi $&$\lambda_\phi$&$y$&$m_f/v_\phi$\\
\midrule
BP1 & $3\times 10^9$ & $-10^{-17}$ & $0.1$ & $4.65\times 10^{-5}$ & $4\times 10^{-5}$\\
BP2 & $6\times 10^4$ & $-10^{-27}$ & $0.1$ & $2.5\times 10^{-8}$ & $5\times 10^{-7}$\\
BP3 & $1.5\times 10^{11}$ & $-3.645\times 10^{-13}$ & $0.1$ & $3\times 10^{-4}$ & $4\times 10^{-4}$\\
\bottomrule[1.2pt]
\end{tabular}
\end{table}

\begin{figure}[!t]
\centering
\includegraphics[width=0.8\textwidth]{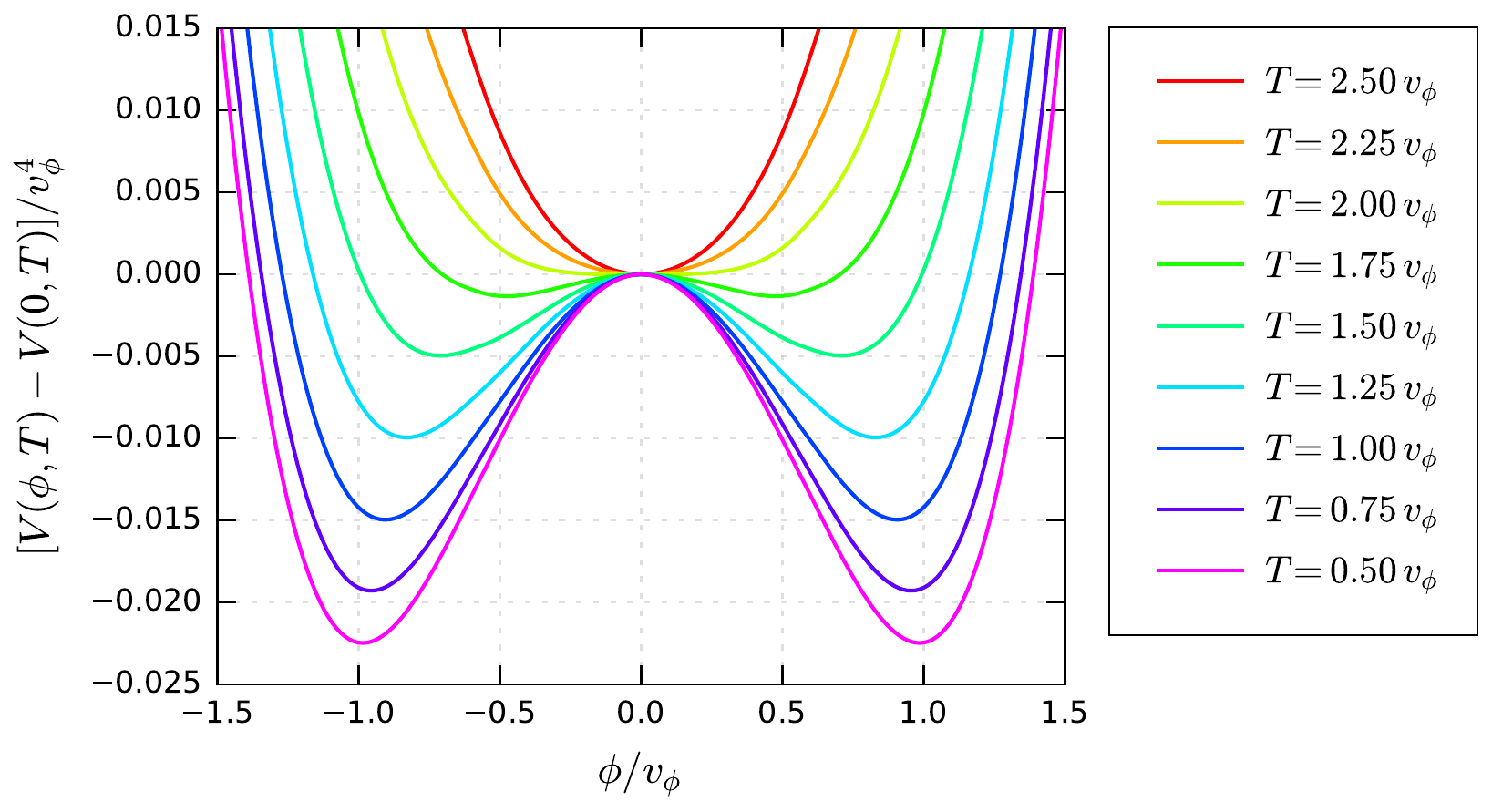}
\caption{Thermally corrected effective potential as functions of $\phi/v_\phi$ with various temperatures for BP1.}\label{fig:potential}
\end{figure}

In Fig.~\ref{fig:potential}, we present the thermally corrected effective potentials as functions of $\phi/v_\phi$ with various temperatures for BP1. The shapes of the potentials for other BPs are similar. The curves are almost identical to the ones obtained with the same $\lambda_\phi$ and $v_\phi$ but without the fermion and the potential bias, and the height of the two local minima of the potential is almost the same. This is because the $\mathbb{Z}_2$-violating couplings $y$ and $\mu_3$ are chosen to be small so that they barely affect the phase transition and that the discrete symmetry is only weakly broken. The tiny potential energy difference $V_{\rm bias}$ between the two local minima plays a crucial role in the evolution of DWs but not in the phase transition.

As shown in Fig.~\ref{fig:potential}, at high temperatures $T\gg v_\phi$, the thermal correction $ V_{\mathrm{T}}$ dominates and the total effective potential $ V $ has a minimum at $ \phi\simeq 0 $, indicating a phase that is approximately symmetric under $ \phi\leftrightarrow -\phi $. 
As the temperature declines,
the relative importance of $ V_0 $ increases and it eventually dominates the effective potential, resulting in two local minima corresponding to a phase where the approximate $\mathbb{Z}_2$ symmetry is spontaneously broken. At low temperatures $T\ll v_\phi$, the locations of the minima tend toward their values at zero temperature, $ \phi_\pm \simeq \pm v_\phi $, where $\phi_-$ is the false vacuum and $\phi_+$ is the true vacuum. 
%Therefore, there is a approximate discrete $\mathbb{Z}_2$ symmetry in $ \mathcal{L} $ under the transforms as $ \phi\rightarrow -\phi $ after phase transition, though $\mathbb{Z}_2$ has been slightly broken primordially. 
According to the curves in Fig.~\ref{fig:potential}, the critical temperature $T_\mathrm{c}$, at which the potential develops two local minima, is $\sim 2 v_\phi$, almost the same as that obtained without the $\mathbb{Z}_2$-violating terms.

\section{Evolution of domain walls}\label{sec:evolution}

%It is an essential precondition for the formation of domain wall. 
%According to the principle of minimum potential energy, the value of $\phi$ will shift from near zero to one of two different local minima.
%{\color{red}(When $\phi_-+\phi_+ \approx 0$ (how to prove $\approx 0$?),
%$$V_{\mathrm{bias}} \approx 
%- \frac{2}{3}\mu_3\phi_+^3 -\frac{1}{8\pi^2}\mu_3\phi_+m_\phi^2( \phi_+)\mathrm{Re}\left[\ln\frac{m_\phi^2( \phi_+)}{\mu_{\mathrm{R}}^2}-1\right]
%+\frac{1}{2\pi^2}y\phi_+M_f^3(\phi_+)\left[ \ln\frac{M_f^2(\phi_+)}{\mu_{\mathrm{R}}^2}-1\right] $$ $$+ V_{\mathrm{T}}(\phi_-)-V_{\mathrm{T}}(\phi_+).$$
%}

In the early Universe after the phase transition, the field values of $\phi$ in causally unconnected regions may lie in different local minima of the effective potential.
A domain wall would appear 
at the boundary separating two regions with different VEVs. 
Because of the potential bias
%% JZ check later
\begin{equation}
V_{\mathrm{bias}}\left(T\right) \equiv V\left(\phi_-,T\right)-V\left(\phi_+,T\right)
\end{equation}
between the false vacuum $\phi_-$ and the true vacuum $ \phi_+ $, the probability $ p_- $ of the scalar field $\phi $ being in $\phi_-$ after the phase transition is different from the probability $ p_+ =1-p_-$ of being in $ \phi_+ $. The ratio of the two is\,\cite{Gelmini:1988sf, Hiramatsu:2010yz}
\begin{equation}
\frac{p_-}{p_+} = \mathrm{e}^{-\Delta F/T},
\end{equation}
where $ \Delta F$ is the difference between the free energies of the two minima, estimated as 
$ \Delta F\simeq V(\phi_-,T)\xi_-^3- V(\phi_+,T)\xi_+^3$, and $ \xi_\pm $ are the correlation lengths of $ \phi $ approximately evaluated by $ \xi_\pm^{-2}\simeq {\partial^2 V( \phi_\pm,T)}/{\partial\phi^2} $ \cite{Kibble:1976sj}. For a large cluster of false vacua to appear in the early Universe, $p_-$ needs to be greater than a critical value $ p_\mathrm{c}\simeq 0.311 $ according to the percolation theory\,\cite{Stauffer:1978kr, Saikawa:2017hiv}. This criterion is easily maintained in this work because the small $\mathbb{Z}_2$-violating couplings keep $\Delta F\ll T$ around the critical temperature. Specifically, we have also numerically verified that all the BPs in Table~\ref{fig:potential} satisfy the condition of $ 0.311 < p_- \leq 0.5 $.

To estimate the evolution of a DW, we numerically calculate the wall tension, i.e., the energy per unit area, by solving the equation of motion\,(EOM) of $\phi$ in the static condition. We will show that within the parameter space of interest, the corrections to the tension by the $\mathbb{Z}_2$-violating terms are negligible.
For a simple evaluation, we consider an instantaneously stable ($ T $-fixed) planar wall invariant in the $yz$ plane so that the DW configuration is described by $ \phi( x,T) $.
The DW interpolates the two different minima at $ x\rightarrow \pm \infty $. The EOM of $ \phi(x,T) $ becomes\,\cite{Vilenkin:2000jqa}
\begin{equation}\label{eq:eq-5}
\frac{\partial^2\phi(x,T)}{\partial x^2}-\frac{\partial V(\phi,T)}{\partial\phi}=0\,,
\end{equation}
%Technically, it is difficult to fix the asymmetric minima at infinity as the boundary conditions of Eq. \eqref{eq:eq-5} in numerical calculation. Moreover, we expect to obtain monotonic solutions corresponding to only one domain wall 
with boundary conditions $ \phi (\pm\infty,T)=\phi_\pm (T) $, where $ \phi_+ (T)$ and $ \phi_- (T) $ are the true and false vacua at temperature $ T $, respectively.
%As for $ x_\pm $,  their absolute values should have been large. Nevertheless, if $ \left|x_\pm\right| $ are too large, the multi-wall or non-smooth solutions will occur via scipy.integrate.solve\_bvp module in python, because Eq. \eqref{eq:eq-5} is a nonlinear partial differential equation and most regions of $ \phi\left( x\right) $ is flat.
%Fortunately, we find that when $ \left|x_\pm\right| $ reach a certain level, the tension of domain wall will barely change, 
For numerical calculation in practice, we set the boundary at $ x_\pm =\pm 20/v_\phi$ to obtain monotonic solutions corresponding to one domain wall.
We have verified that the obtained DW tension is rather insensitive to this choice.

The energy density of the solution,\footnote{The one-loop corrections to the kinetic term of $\phi$ are incorporated into the field strength renormalization constant $Z_\phi$ through the renormalization condition that sets the kinetic term to its canonical form.
%which is universal for the kinetic and potential terms involving $\phi$.
Therefore, the one-loop corrections merely induce a small rescaling of $\phi$, whose effect on $\rho_\mathrm{DW}$ is negligible here.}
\begin{equation}
\rho_{\mathrm{DW}}= \frac{1}{2}\,| \nabla \phi|^2 + V( \phi,T)-V_{\mathrm{min}}( \phi,T)\,,
\end{equation}
has a peak at $x_{\mathrm{peak}}$, 
%between $ x_\pm $, denoted as $ \rho_{\mathrm{DW}}\left( x_{\mathrm{peak}},T\right) $, since the place around $ x_{\mathrm{peak}} $ is 
which is taken as the center of the DW.
The thickness of DW is approximately $\delta\sim\left(\sqrt{2\lambda_\phi}v_\phi\right)^{-1}$ \cite{Vilenkin:2000jqa}, and $\delta$ is just a small portion of $\left(x_+-x_-\right)$. Thus, between $x_-$ and $x_+$, there is a thin region of the DW and large regions of the two vacua. To avoid counting the false vacuum energy when calculating the tension of the DW, we introduce two boundary points $\delta_\pm$ between $x_\pm$ and $x_{\mathrm{peak}}$ to isolate the DW. Specifically, we define $\delta_\pm$ with the condition
\begin{equation}
\frac{\rho_{\mathrm{DW}}(x_{\mathrm{peak}})-\rho_{\mathrm{DW}}( \delta_\pm) }{\rho_{\mathrm{DW}}(x_{\mathrm{peak}}) -\rho_{\mathrm{DW}}(x_\pm)}=0.99 \quad( x_-<\delta_-<x_{\mathrm{peak}}<\delta_+<x_+)\,.  
\end{equation}
Accordingly, the DW tension is evaluated as
\begin{equation}
\sigma_{\mathrm{DW}}(T) =\int_{\delta_-}^{\delta_+}\rho_{\mathrm{DW}}(x,T)\,\mathrm{d}x\,.
\end{equation}

\begin{figure}[!t]
\centering
\includegraphics[width=0.5\textwidth]{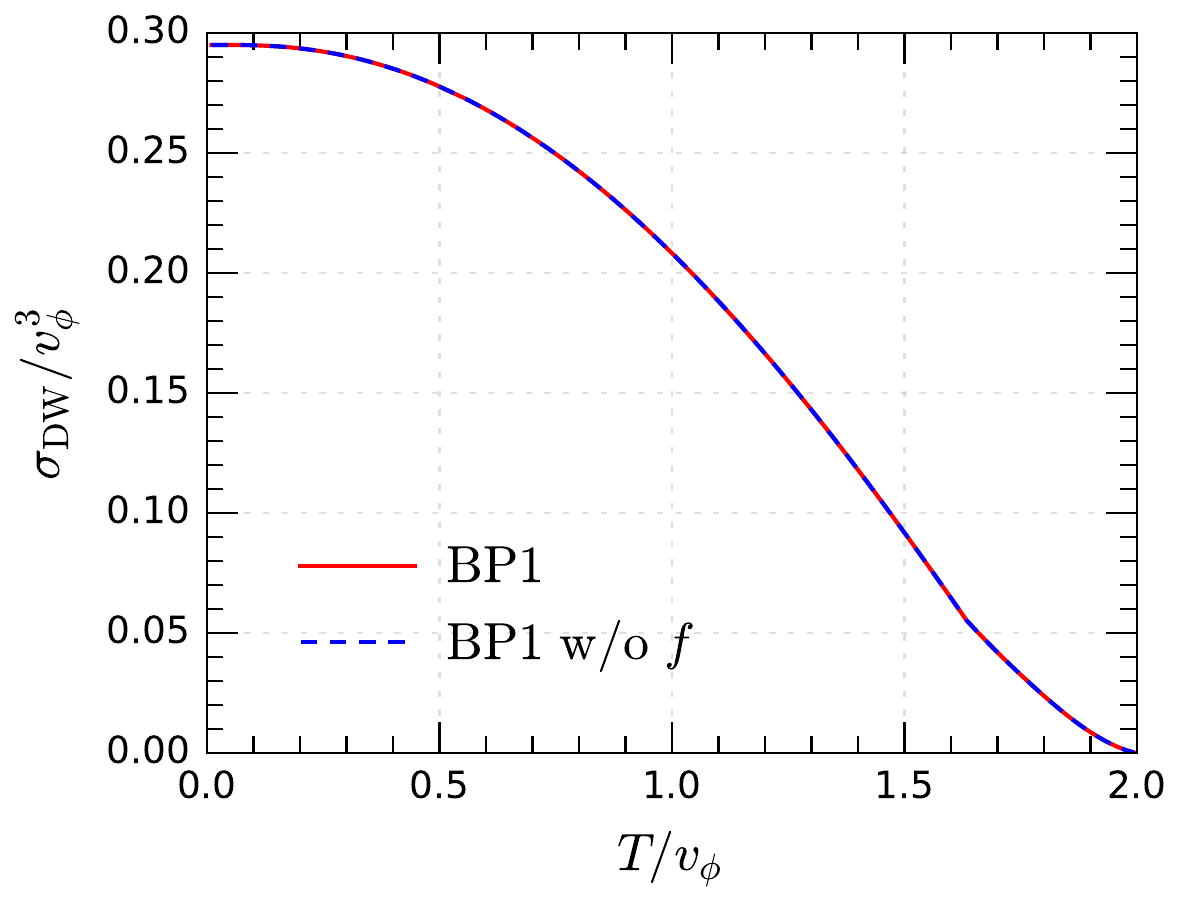}
\vspace{-0.7em}\caption{DW tension varying with temperature. The solid red line and dashed blue line correspond to BP1 with and without the fermion $f$, respectively.}\label{fig:tension}
\end{figure}

Figure \ref{fig:tension} shows the obtained DW tension $\sigma_{\mathrm{DW}}$ as a function of the temperature for BP1 with and without coupling to the fermion.
The $\mathbb{Z}_2$-violating couplings $y$ and $\mu_3$ have little influence on $ \sigma_{\mathrm{DW}}$ because of their tiny values. After the phase transition, $\sigma_{\mathrm{DW}}$ grows from zero toward $ 2\sqrt{2\lambda_\phi}v_\phi^3/3$ at low temperatures, resembling the evolution of a DW with a $\mathbb{Z}_2$-symmetric potential \cite{Vilenkin:2000jqa}.
The tensions obtained for other BPs have similar features because of the small $y$ and $\mu_3$.

Once formed, the DW network evolves toward the scaling regime, in which various length scales, such as the typical radius of curvature $ R_{\mathrm{DW}} $ of the walls and the average distance between the walls, are comparable to the Hubble radius.
As a result, one has $ R_{\mathrm{DW}}\simeq t/\mathcal{A} \sim (\mathcal{A} H)^{-1}$ \cite{Press:1989yh, Hindmarsh:1996xv, Hindmarsh:2002bq, Garagounis:2002kt, Hiramatsu:2010yz, Hiramatsu:2013qaa}, and the energy density of the DWs can be expressed as
\begin{equation}
\rho_\mathrm{DW} = \frac{\mathcal{A}\,\sigma_\mathrm{DW}}{t}
\,,
\end{equation}
where the parameter $ \mathcal{A}\simeq 0.8 $ is determined by field theoretic simulation \cite{Hiramatsu:2013qaa} and is almost constant in the scaling regime.
We assume that the DW network evolves and eventually collapses in the radiation-dominated era, so the Hubble rate is given by \cite{Kolb:1990vq}
\begin{equation}\label{eq:H(T)}
H(T)=\sqrt{\frac{4\pi^3Gg_*(T)}{45}}\,T^2\,,
\end{equation}
with $G \simeq 6.7\times10^{-39}\,\mathrm{GeV}^{-2}$ \cite{ParticleDataGroup:2024cfk} the Newtonian gravitational constant and $g_*$ the number of relativistic degrees of freedom \cite{Kolb:1990vq, Baumann:2022mni}.

Multiple counteracting forces contribute to the evolution of a DW. The expansion of a DW is driven by the tension force per unit area,
\begin{equation}
p_{\mathrm{T}}\sim \frac{\sigma_{\mathrm{DW}}}{R_{\mathrm{DW}}}
\simeq \rho_{\mathrm{DW}}\,.
\label{eq:tension}
\end{equation}
When the potential bias is small, $\sigma_{\mathrm{DW}}$ is mainly determined by the $\mathbb{Z}_2$-symmetric part of the scalar potential and is nearly constant at low temperatures, so that $ p_\mathrm{T} \propto T^2 $ in the scaling regime for $T \ll v_\phi$ during the radiation-dominated era. 
The potential bias $ V_{\mathrm{bias}} $ 
provides a pressure, 
\begin{equation}
p_{\mathrm{V}}\sim V_{\mathrm{bias}}\,,
\label{eq:p_V}
\end{equation}
that collapses the wall.
The friction force per unit area $F_f$ exerted by the particles in the ambient environment, which are the $f$ fermions in this work, opposes the motion of the wall\,\cite{Vilenkin:2000jqa, Nakayama:2016gxi, Blasi:2022ayo, Blasi:2023sej}.
%Finally, domain walls will vanish while $ p_{\mathrm{T}} < p_{\mathrm{V}} $ due to the cosmic expansion.
The DW network starts to collapse at the annihilation temperature 
$T_{\mathrm{ann}} $ when
\begin{equation}\label{eq:ann_eq}
p_{\mathrm{V}}+F_f\simeq p_{\mathrm{T}}\,.
\end{equation}

The interaction between the DW and the $f$ particles in the thermal bath induces friction on the wall as it moves in the plasma. To estimate the friction, we consider the EOM of the fermion $f $,
\begin{equation}\label{eq:fEOM}
\left[ \partial^2+\left(m_f+y\phi\right)^2\right]f(\mathbf{x},t) = 0.
\end{equation}
We begin with estimating the relative motion between a Dirac fermion $f $ and a DW in the DW rest frame with the DW configuration $\phi( x,T)$ discussed in Eq.~\eqref{eq:eq-5}. The fermion wave function has the form
$f(\mathbf{x},t) = g(x)\,\mathrm{e}^{\mathrm{i}\left(-\omega t + p_y y + p_z z \right)} $.
%, where $ \omega $ is the energy of $ f $, and $ p_y $, $ p_z $ are the $ y$, $z$-components of momentum. 
Substituting it into Eq.~\eqref{eq:fEOM}, we obtain
\begin{equation}
\frac{\mathrm{d}^2g(x)}{\mathrm{d}x^2}+p_x^2g(x)-U[ \phi( x)] g(x)=0\,,
\end{equation}
where $p_x^2=\omega^2-p_y^2-p_z^2-m^2 $ and $U(\phi)=2ym_f\phi+y^2\phi^2$.  
%For $ m_f \gg |yv_\phi|$, 
For $p_x\ll \delta^{-1}$ with $\delta\sim\left(\sqrt{2\lambda_\phi}v_\phi\right)^{-1}$ the thickness of the wall\,\cite{Vilenkin:2000jqa}, this collision problem can be approximated by the classic one-dimensional scattering of a free particle by a step potential, with $\phi(x)$ approximated as
\begin{equation}
\phi(x)\simeq \left\lbrace \begin{array}{ll}
\phi_-&\left(x<x_{\mathrm{peak}}\right),\\
\phi_+&\left(x>x_{\mathrm{peak}}\right),
\end{array}\right.
\end{equation}
recalling that we have chosen the coordinate so that the true\,(false) vacuum lies at $x>0$\,($x<0$).
For a fermion traveling from a vacuum $\phi_i$ to another vacuum $\phi_j$, the probability of it being reflected by the DW is
\begin{equation}
R_{\phi_i\rightarrow\phi_j}\simeq\left\lbrace 
\begin{array}{ll}
\dfrac{\left[U(\phi_i)-U(\phi_j)\right]^2}{\left[\sqrt{p_x^2-U( \phi_i)}+\sqrt{p_x^2-U( \phi_j)}\right]^4}, &\quad p_x^2\geq \mathrm{max}\left\lbrace U(\phi_i),U(\phi_j)\right\rbrace ,\\[2em]
1, &\quad U(\phi_i)\leq p_x^2<U( \phi_j),
\end{array}\right.
\label{eq:refl_prob}
\end{equation}
where $\phi_{i/j}= \phi_{+/-}$ or $ \phi_{-/+} $. Note that values of $ p_x $ outside the ranges quoted in Eq.~\eqref{eq:refl_prob} correspond to bound state solutions and are irrelevant to the calculation of friction.
The actual reflectivity of $ f $ is generally smaller than that obtained with this approximation because the DW configuration $ \phi(x) $ is gentler than a steep step, and so is $ U[\phi(x)] $.

%We assume that plasma particles are in thermal equilibrium state and take the random thermal motion, and 
DWs expand after formation with an average velocity $v_{\mathrm{DW}}$. 
%along the $x$-axis in a background reference frame
To estimate the friction force, we consider a DW moving along the $x$ axis at velocity $ v_{\mathrm{DW}} $. According to the convention above,
$ v_{\mathrm{DW}} $ is positive when the DW moves from the false vacuum to the true one. 
For a fermion $f$ in the thermal bath with momentum $p$, its velocity relative to the DW is $ ( p_x -\omega v_{\mathrm{DW}})/(\omega-p_xv_{\mathrm{DW}})$, and the momentum transfer in a collision with the wall is $ -2( p_x -\omega v_{\mathrm{DW}})/(1-v_{\mathrm{DW}}^2)$. 
%The number of states of $ f $ between $\vec{p}$ and $\vec{p}+\mathrm{d}\vec{p}$ is $$\mathrm{d}n=\frac{g_f}{\left(2\pi\right)^3}\frac{1}{e^{\omega/T}+1}\,\mathrm{d}^3\vec{p},$$
%where $g_f=4$ is the degrees of freedom of $ f $.
Therefore, the friction force per unit area %with special-relativity effect
exerted on the DW  
is\,\cite{Kolb:1990vq, Nakayama:2016gxi, Vilenkin:2000jqa, Kibble:1976sj}
\begin{equation}\label{eq:friction}
F_f=\frac{2}{\pi^2} \frac{1}{1-v_{\mathrm{DW}}^2}\int_0^{+\infty}\int_{-\infty}^{+\infty}R(p_x)\,\frac{( p_x -\omega v_{\mathrm{DW}})^2}{\omega-p_xv_{\mathrm{DW}}} \frac{1}{\mathrm{e}^{\omega/T}+1}\, p_\perp \,\mathrm{d}p_x\,\mathrm{d}p_\perp,
\end{equation}
where $ p_\perp \equiv \sqrt{p_y^2+p_z^2}$, 
and the reflection probability $ R(p_x) $ is defined as
\begin{equation}
R(p_x ) =\left\lbrace \begin{array}{ll}
R_{\phi_+\rightarrow\phi_-}&\left(p_x<p_0\right),\\
-R_{\phi_-\rightarrow\phi_+}&\left(p_x> p_0\right),
\end{array}\right.
\end{equation}
where $p_0=v_{\mathrm{DW}}\sqrt{(p_\perp^2+m^2)/(1-v_{\mathrm{DW}}^2)}$, and $p_x=p_0 $ indicates that the fermion is at rest with respect to the wall.

Before proceeding to the numerical results, we present qualitative%
\footnote{In the following qualitative estimates, we ignore numerical factors and only consider the scaling behavior.}
estimates for the more analytically calculable regime of 
$v_\phi\gg T\gg m_f\gg |y v_\phi|$, where the phase transition has settled while the temperature remains high enough to keep the fermions in the thermal bath. The limit $m_f\gg |y v_\phi|$ allows us to treat $\mathbb{Z}_2$-violating effects as perturbations. The results for the limit $m_f\ll |y v_\phi|$ share the same qualitative behavior to be discussed below.

In this limit with a sufficient small $\mu_3$, the dominant thermal correction to the effective potential comes from the fermions in the thermal bath, so that the thermal contribution to $V_\mathrm{bias}$ is roughly  $\Delta V_\mathrm{T}\equiv V_\mathrm{T}(\phi_-,T)-V_\mathrm{T}(\phi_+,T)\sim y v_\phi m_f T^2 $. The friction can be estimated as
$F_{f}\sim n_{f}\Delta p\,R\sim T^{4}v_\mathrm{DW}\,R$,
where $\Delta p$ is the momentum exchange and $n_f$ is the number density of the fermions. The scaling of the reflection probability can be estimated according to Eq.~\eqref{eq:refl_prob}, $R\sim (y v_\phi)^2 m_f^2/\omega^4$, with the typical particle energy $\omega\sim T$. Therefore, the friction is negligible in this limit compared to the pressure, $F_f\ll \Delta V_\mathrm{T} < V_\mathrm{bias} \sim p_\mathrm{V}$. The DWs thus collapse at the time when 
$p_\mathrm{V}/p_\mathrm{T} \gtrsim 1$.

If $V_\text{bias}(T=0) \gg \Delta V_\mathrm{T}$, 
$V_\mathrm{bias}$ is dominated by the asymmetric terms in the zero-temperature scalar potential, which are independent of the temperature. This reduces to the conventional scenario studied extensively in the literature without the background fermionic thermal bath. On the other hand, 
if $V_\text{bias}(T=0) \ll \Delta V_\mathrm{T}$ before the DW network collapses, $V_\mathrm{bias}$ and thus the pressure is mainly determined by the thermal correction by the fermions, which translates to $p_\mathrm{V}/p_\mathrm{T}\sim \Delta V_\mathrm{T}/p_\mathrm{T} \sim |y v_\phi| m_f M_\mathrm{P}/ \sigma_\mathrm{DW}$ according to
Eqs.~\eqref{eq:H(T)}--\eqref{eq:p_V}, where $M_\mathrm{P}$ is the Planck mass. 
In this case, $p_\mathrm{V}/p_\mathrm{T}$ is almost temperature independent for $m_f\ll T\ll v_\phi$.

If $|y v_\phi| m_f M_\mathrm{P} \gg \sigma_\mathrm{DW}$, the thermal correction renders the DWs unstable right after they are formed, and they may not evolve into the scaling regime. Therefore, it is important to examine the size of the thermal correction to $V_\mathrm{bias}$ for the evolution of the DWs. If $|y v_\phi| m_f M_\mathrm{P} \ll \sigma_\mathrm{DW}$, the thermal effect alone is too weak to make the DWs collapse, and the evolution is again determined by the temperature-independent part $V_\text{bias}(T=0)$. For the intermediate regime between the two extremes, the thermal correction is small enough to keep the DW network metastable so that it evolves into the scaling regime, while large enough to compete with $V_\text{bias}(T=0)$ to affect the late time evolution and its accompanying phenomenology. However, it is no longer easy to estimate the results analytically. The BPs fall within this regime and we present their numerical results in the following.  

\begin{figure}[!t]
\centering
\includegraphics[width=0.5\textwidth]{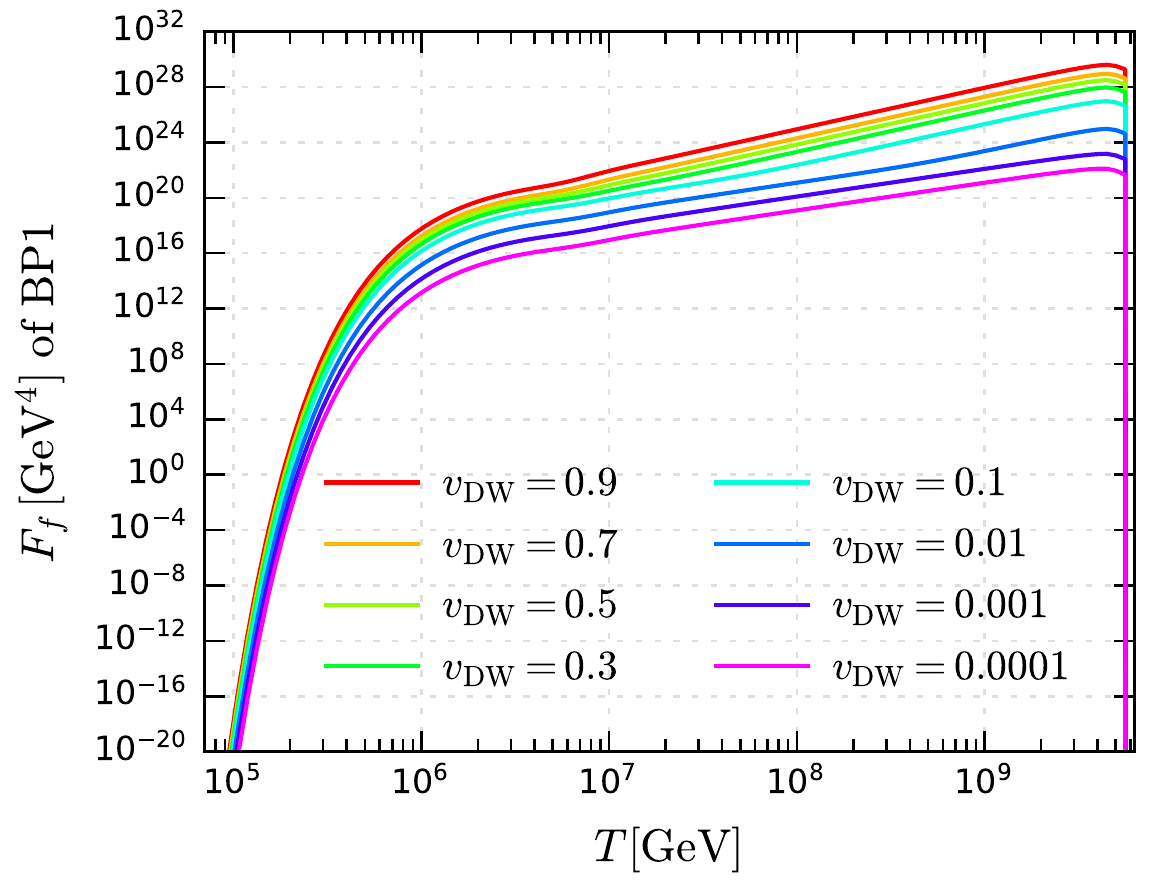}
\caption{Friction force per unit area $F_f$ exerted on the DW as a function of the temperature with various wall velocities $ v_\mathrm{DW} $ for BP1.}\label{fig:friction}
\end{figure}

In Fig.~\ref{fig:friction}, we plot the friction $ F_f $ on the DW as a function of the temperature with various wall velocities $ v_\mathrm{DW} $ for BP1.
The frictions evaluated for other BPs share a similar behavior qualitatively. 
At high temperatures close to the critical temperature of the phase transition, the calculated friction falls drastically because of the shallower potential difference between the true and false vacua, reducing the reflection probability of the incoming fermion to the wall. 
At low temperatures, $ F_f $ decreases exponentially due to the Boltzmann suppression when the temperature becomes comparable to the fermion mass. %Therefore, if $ T_\mathrm{ann} $ is much lower than the phase transition temperature, $ F_f $ usually has no effect on the annihilation process, and $ F_f \ll p_\mathrm{V}$ usually. 
For smaller $v_\mathrm{DW}$, the friction $F_f$ decreases slowly due to the smaller momentum transfer per collision between the wall and the fermion. 
For simplicity, we will assume the typical velocity $ v_{\mathrm{DW}} \simeq 0.3$ \cite{Kawano:1989mw, Avelino:2005kn} for the rest of this work.

As the Universe expands, the energy density of the DWs in the scaling regime dilutes slower than those of matter and radiation. If the DW network collapses late, it could dominate the energy density of the Universe at the temperature $ T_\mathrm{dom} $ corresponding to\,\cite{Gelmini:1988sf, Hiramatsu:2010yz, Hiramatsu:2013qaa, Nakayama:2016gxi, Saikawa:2017hiv}
\begin{equation}
p_\mathrm{T} \sim \mathcal{A}\,\sigma_{\mathrm{DW}}H\sim\rho_{\mathrm{DW}}\simeq\rho_{\mathrm{c}}=\frac{3H^2}{8\pi G},
\end{equation}
where $\rho_\mathrm{c}$ is the critical density.
For a consistent cosmological history, the DWs need to collapse before they overclose the Universe. This sets the condition  $T_{\mathrm{ann}} > T_\mathrm{dom}$, which is equivalent to $p_\mathrm{T} \lesssim \rho_\mathrm{c}$ at $T_{\mathrm{ann}}$.

\begin{figure}[tb]
\centering
\subfigure[BP1\label{fig:ann:a}]{\includegraphics[width=0.485\textwidth]{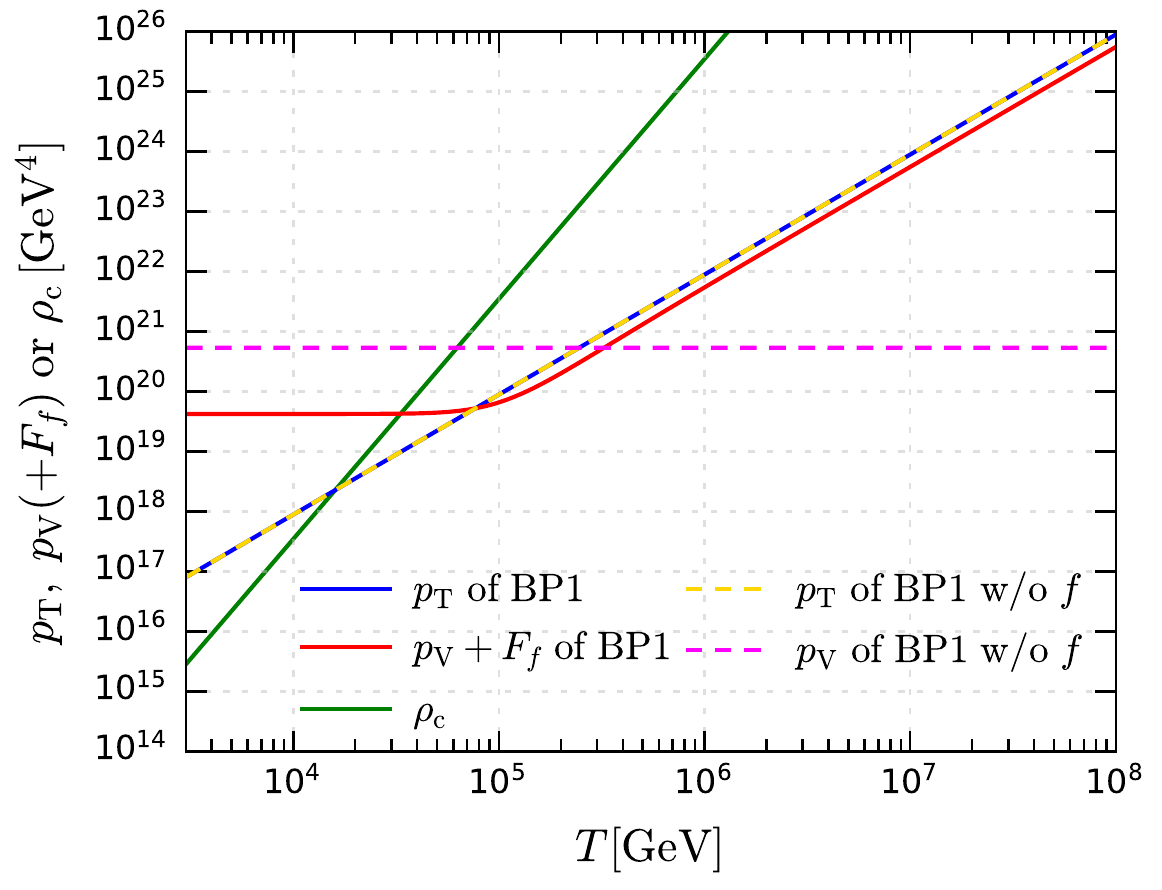}}
\subfigure[BP2\label{fig:ann:b}]{\includegraphics[width=0.5\textwidth]{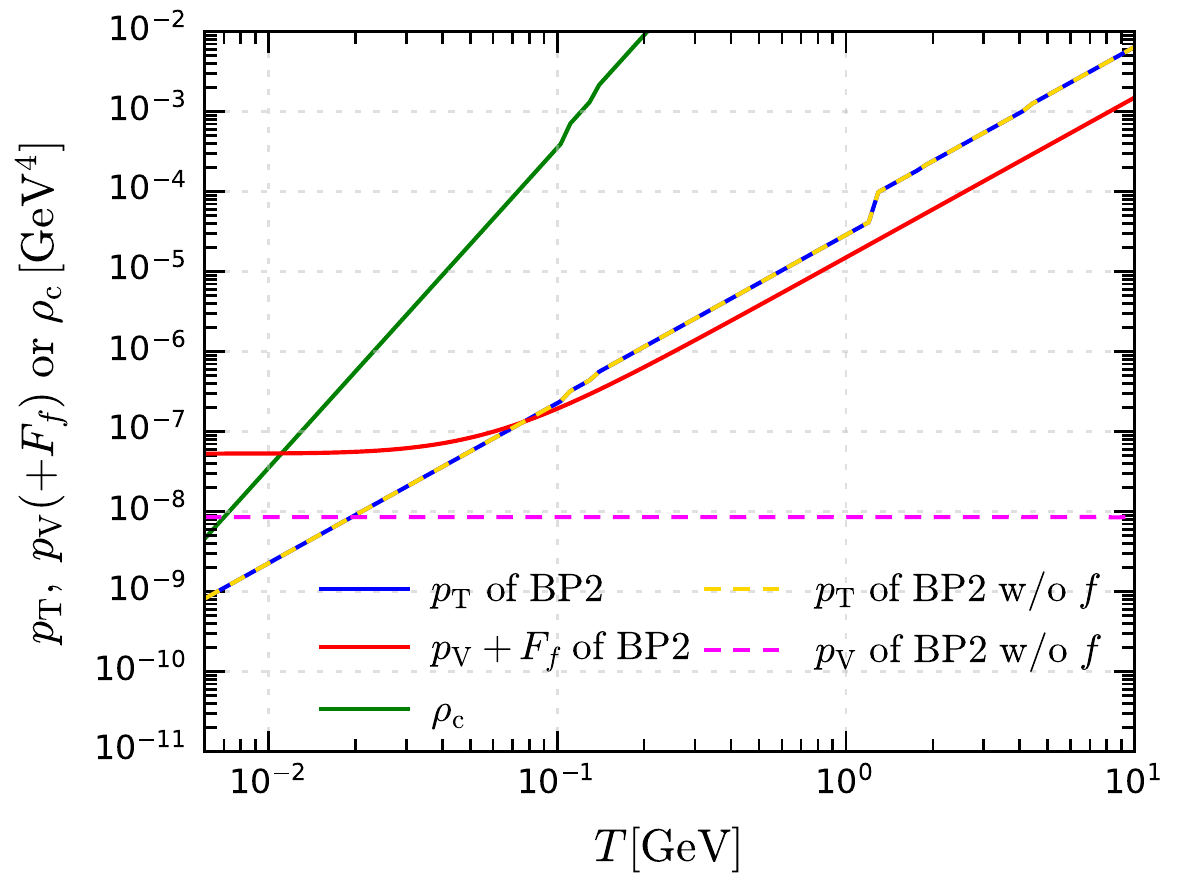}}
\subfigure[BP3\label{fig:ann:c}]{\includegraphics[width=0.52\textwidth]{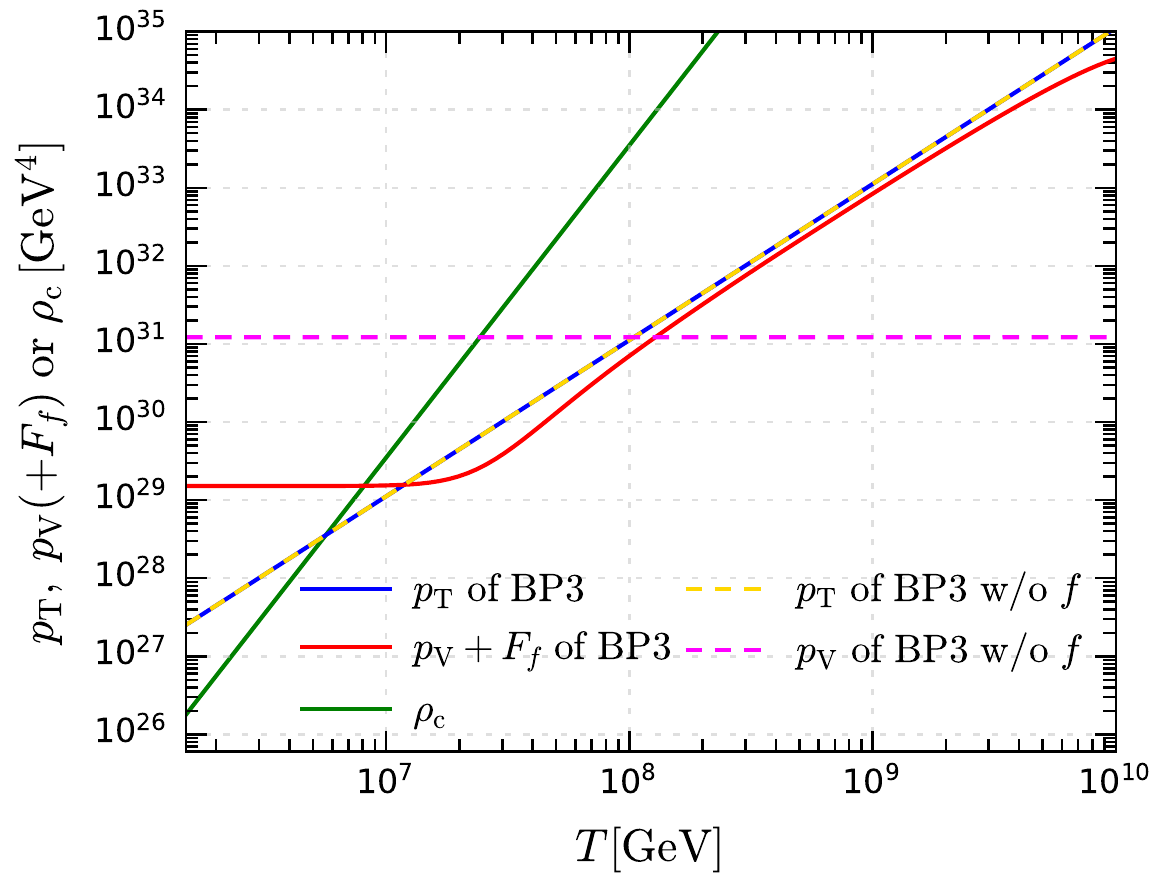}}
\caption{$ p_\mathrm{T} $, $p_\mathrm{V}+F_f $, $p_\mathrm{V}$, and $ \rho_\mathrm{c} $ as functions of the temperature $T$ for (a)~BP1, (b)~BP2, and (c)~BP3. The solid (dashed) lines correspond to the result with (without) the fermion $f$.}\label{fig:ann}
\end{figure}

In Fig.~\ref{fig:ann}, we summarize all the previous considerations on the evolution and the collapse of DWs by plotting $ p_\mathrm{T} $, 
$ p_\mathrm{V}(+F_f) $, and $ \rho_\mathrm{c} $ as functions of the temperature for the three BPs.
For each BP, the two curves for the tension force $ p_\mathrm{T} $ with and without the fermion overlap with each other, because the DW tensions are almost unaltered by the small Yukawa coupling, as shown in Fig.~\ref{fig:tension}.
After the phase transition, the $ p_\mathrm{T} $ curves in the three BPs decrease with the temperature as $ T^2 $ during the radiation-dominated era according to Eq.~\eqref{eq:tension}. The steplike features at $T \sim \si{GeV}$ and $T \sim 10^{-1}~\si{GeV}$ for BP2 are caused by the thresholds in the effective degrees of freedom $ g_*\left(T\right) $.

In the temperature ranges shown in Fig.~\ref{fig:ann}, the calculated friction $ F_f $ with $ v_\mathrm{DW} =0.3$ for the three BPs are always negligible, compared to $p_\mathrm{V}$.
Since the $\mu_3$ values in the BPs are extremely small, the thermal correction to $ p_\mathrm{V} $ induced by $\mu_3$ is insignificant, compared to the thermal correction from the Yukawa coupling, which renders $ p_\mathrm{V} \propto T^2$ at high temperatures.
Therefore, all the curves of $ p_\mathrm{V}+F_f $ in the three BPs are basically parallel to $ p_\mathrm{T} $ at high temperatures. At low temperatures $T\lesssim |M_f(v_\phi)|$, the potential bias induced by thermal corrections becomes negligible and the curves of $ p_\mathrm{V} \left(+F_f\right)$ become flat. 
Here, the differences between BP$n$ and BP$n$ w/o $f$ are caused by the Coleman-Weinberg effective potential $ V_\mathrm{CW}$.

\begin{table}[!t]
\caption{Values of $ T_\mathrm{ann} $, $f_\mathrm{peak}$, and $\Omega_\mathrm{GW}(f_\mathrm{peak}) h^2$ for the BPs.}\label{tab:result}
\belowrulesep=0pt
\aboverulesep=0pt
\centering
\renewcommand{\arraystretch}{1.5}
\begin{tabular}{@{\hspace{1.5em}}c@{\hspace{4em}}c@{\hspace{4em}}c@{\hspace{4em}}c@{\hspace{1.5em}}}
\toprule[1.2pt]
&$ T_\mathrm{ann}\,(\mathrm{GeV})$&$f_\mathrm{peak}\,(\mathrm{Hz})$&$\Omega_\mathrm{GW}(f_\mathrm{peak}) h^2$\\
\midrule
BP1& $ 7.67\times 10^4 $& $1.28\times 10^{-2}$& $2.49\times 10^{-9}$\\
BP1 w/o $ f $&$2.45\times 10^5$&$4.08\times 10^{-2}$&$2.39\times 10^{-11}$\\
\midrule
BP2&$ 7.62\times 10^{-2}$&$8.66\times 10^{-9}$&$3.49\times 10^{-12}$\\
BP2 w/o $ f $&$1.94\times 10^{-2}$&$2.20\times 10^{-9}$&$8.31\times 10^{-10}$\\
\midrule
BP3&$ 1.19\times 10^7$&$1.97$&$6.82\times 10^{-8}$\\
BP3 w/o $ f $&$1.04\times 10^8$&$17.4$&$1.14\times 10^{-11}$\\
\bottomrule[1.2pt]
\end{tabular}
\end{table}

The DW annihilation temperature $T_\mathrm{ann}$ is determined by the intersection between the curves of 
$p_\mathrm{V}(+F_f)$ and $p_\mathrm{T}$, and the obtained $T_\mathrm{ann}$ values for the BPs are given in Table~\ref{tab:result}.
For the results with the fermion $f$ shown in Fig.~\ref{fig:ann}, the $p_\mathrm{T}$ curves intersect with the $p_\mathrm{V}+F_f$ curves at which the $p_\mathrm{V}+F_f$ curves start to rise with $T$, indicating that the thermal corrections are getting important.
Since the $p_\mathrm{T}$ and  $p_\mathrm{V}+F_f$  curves are parallel to each other at high temperatures, their intersection points, and thus $T_\mathrm{ann}$, are rather sensitive to increasing values of $|y|$ and $|m_f|$.
For BP1 and BP3, the DWs collapse later with lower $T_\mathrm{ann}$ than their counterpart without the fermion, affected mainly by the differences in the  
Coleman-Weinberg effective potential 
$V_\mathrm{CW}$. It is the converse for BP$2$.
%Whilst for BP$2$, the correction by $V_\mathrm{CW}$ is almost negligible. The earlier collapse time of the DWs in BP2 relative to that without the fermion is driven by the thermal correction from the fermion.

Finally, if the DWs do not collapse, they dominate the early Universe at the temperature $T_\mathrm{dom}$, which is determined by the intersection between the curves of $p_\mathrm{T}$ and $\rho_\mathrm{c}$. 
In all the BPs, the annihilation temperatures $ T_{\mathrm{ann}}$ are higher than both $ T_{\mathrm{dom}} $ and the big bang nucleosynthesis temperature $\sim 1\,\mathrm{MeV} $~\cite{ParticleDataGroup:2024cfk, Kolb:1990vq}, indicating a consistent cosmological history.

\section{Spectrum of gravitational waves}\label{spectrum}

The DW network undergoes dynamical evolution from its initial formation to eventual collapse, and a portion of its energy density would be radiated as GWs.
A rough estimate of the energy density of the emitted GWs can be obtained by the quadrupole approximation, where the quadrupole moment of the DWs is evaluated by $ Q_{ij} \sim \mathcal{A} \sigma_{\mathrm{DW}}t^4$ in the scaling regime. Thus, the power of gravitational radiation is estimated as $
P\sim G \dddot{Q}_{ij}(t)\dddot{Q}_{ij}(t)
\sim G\mathcal{A}^2\sigma_{\mathrm{DW}}^2t^2$,
and the GW energy density is given by $\rho_{\mathrm{GW}}\sim Pt/{t^3}\sim G\mathcal{A}^2\sigma_{\mathrm{DW}}^2$ \cite{Hiramatsu:2012sc, Maggiore:2007ulw}.
Although such an estimate may not be accurate \cite{Kitajima:2023cek,Kitajima:2023kzu}, it reveals that the GW energy density generated by DWs is basically proportional to $G\mathcal{A}^2\sigma_{\mathrm{DW}}^2$.

The SGWB produced by collapsing 
DWs has been evaluated via numerical simulations \cite{Hiramatsu:2010yz, Hiramatsu:2013qaa, Saikawa:2017hiv, Kawasaki:2011vv, Ferreira:2022zzo, Caprini:2019egz}.
The frequency spectrum of the SGWB is commonly represented by a dimensionless quantity
\begin{equation}
\Omega_\mathrm{GW}(f) \equiv \frac{1}{\rho_\mathrm{c}} \frac{\mathrm{d}\rho_\mathrm{GW}}{\mathrm{d}\ln f},
\end{equation}
where $f$ is the GW frequency.
Previous research indicates that the majority of GWs emitted from DWs is at a peak frequency $ f_{\mathrm{peak}}(T_\mathrm{ann}) \sim H(T_\mathrm{ann})$, corresponding to the time when DWs collapse at the annihilation temperature $ T_{\mathrm{ann}} $. Taking into account the cosmological redshift, the peak frequency today becomes \cite{Hiramatsu:2010yz, Hiramatsu:2013qaa}
\begin{equation}\label{eq:f_peak}
f_{\mathrm{peak}}(T_0)
=\frac{a(T_{\mathrm{ann}})H(T_{\mathrm{ann}})}{a(T_0)}
=\left[ \frac{g_{*S}(T_0)}{g_{*S}(T_{\mathrm{ann}})} \right]^{1/3} \frac{T_0 H(T_{\mathrm{ann}})}{T_{\mathrm{ann}}} ,
\end{equation}
where $a(T)$ is the scale factor at temperature $T$, $ T_0 = 2.7255~\si{K}$ 
is the present temperature of the Universe, and $ g_{*S}(T) $ is the number of entropic relativistic degrees of freedom \cite{Kolb:1990vq, Baumann:2022mni}.
We set $ g_{*S}(T_{\mathrm{ann}}\gtrsim 100~\si{GeV}) \simeq 100$ and $g_{*S}(T_0)=3.91$.

The peak amplitude of $ \Omega_{\mathrm{GW}}(f)$ at the annihilation temperature is \cite{Hiramatsu:2010yz, Hiramatsu:2013qaa}
\begin{equation}\label{eq:OmegaGW_peak_Tann}
\Omega_{\mathrm{GW}}(f_{\mathrm{peak}})\big|_{T_{\mathrm{ann}}}
= \frac{8\pi\tilde{\epsilon}_{\mathrm{GW}}G^2\mathcal{A}^2\sigma_{\mathrm{DW}}^2(T_{\mathrm{ann}})}{3H^2(T_{\mathrm{ann}})},
\end{equation}
where $ \tilde{\epsilon}_{\mathrm{GW}}= 0.7 $ is a factor determined by numerical simulations \cite{Hiramatsu:2013qaa}.
Taking account of the cosmic expansion, the peak amplitude at the present can be expressed as \cite{Saikawa:2017hiv}
\begin{equation}
\Omega_{\mathrm{GW}}(f_{\mathrm{peak}})\big|_{T_0}
=\Omega_{\mathrm{rad}}(T_0)\,\frac{g_*(T_{\mathrm{ann}})}{g_*(T_0)}\left[ \frac{g_{*S}(T_0)}{g_{*S}(T_{\mathrm{ann}})} \right]^{\frac{4}{3}} \Omega_{\mathrm{GW}}(f_{\mathrm{peak}})\big|_{T_{\mathrm{ann}}},
\end{equation}
where $ \Omega_{\mathrm{rad}}\left(T_0\right)=1.68\times 5.38\times 10^{-5} $ 
is the fraction of the radiation energy density today \cite{ParticleDataGroup:2024cfk} and we adopt $ g_*(T_{\mathrm{ann}}\gtrsim 100 ~\si{GeV}) \simeq 100$ and $ g_*(T_0)=3.36$.
Furthermore, causality implies $\Omega_\mathrm{GW} \propto f^3$ for $f < f_{\mathrm{peak}}$\,\cite{Caprini:2009fx,Cai:2019cdl}, while numerical simulations suggest $\Omega_\mathrm{GW} \propto f^{-1}$ for $f > f_{\mathrm{peak}}$\,\cite{Hiramatsu:2013qaa}.\footnote{Numerical simulations in Refs.~\cite{Ferreira:2023jbu,Dankovsky:2024zvs} yield slightly different spectral indices for $f > f_{\mathrm{peak}}$.}
Therefore, the present SGWB spectrum can be formulated as
\begin{equation}
\Omega_{\mathrm{GW}}(f) = \Omega_{\mathrm{GW}}(f_{\mathrm{peak}})\big|_{T_0} \times\left\lbrace 
\begin{array}{ll}
\left(\dfrac{f}{f_{\mathrm{peak}}}\right)^3, & \quad f\leq f_{\mathrm{peak}},\\[1em]
\left(\dfrac{f}{f_{\mathrm{peak}}}\right)^{-1}, & \quad f>f_{\mathrm{peak}}.
\end{array}\right.
\end{equation}
When comparing with GW experiments, the SGWB spectrum is typically expressed as $\Omega_\mathrm{GW} h^2$, where $h = 0.674$ \cite{ParticleDataGroup:2024cfk} is the Hubble constant in units of $100~\si{km~s^{-1}~Mpc^{-1}}$.

\begin{figure}[!t]
\centering
\includegraphics[width=0.8\textwidth]{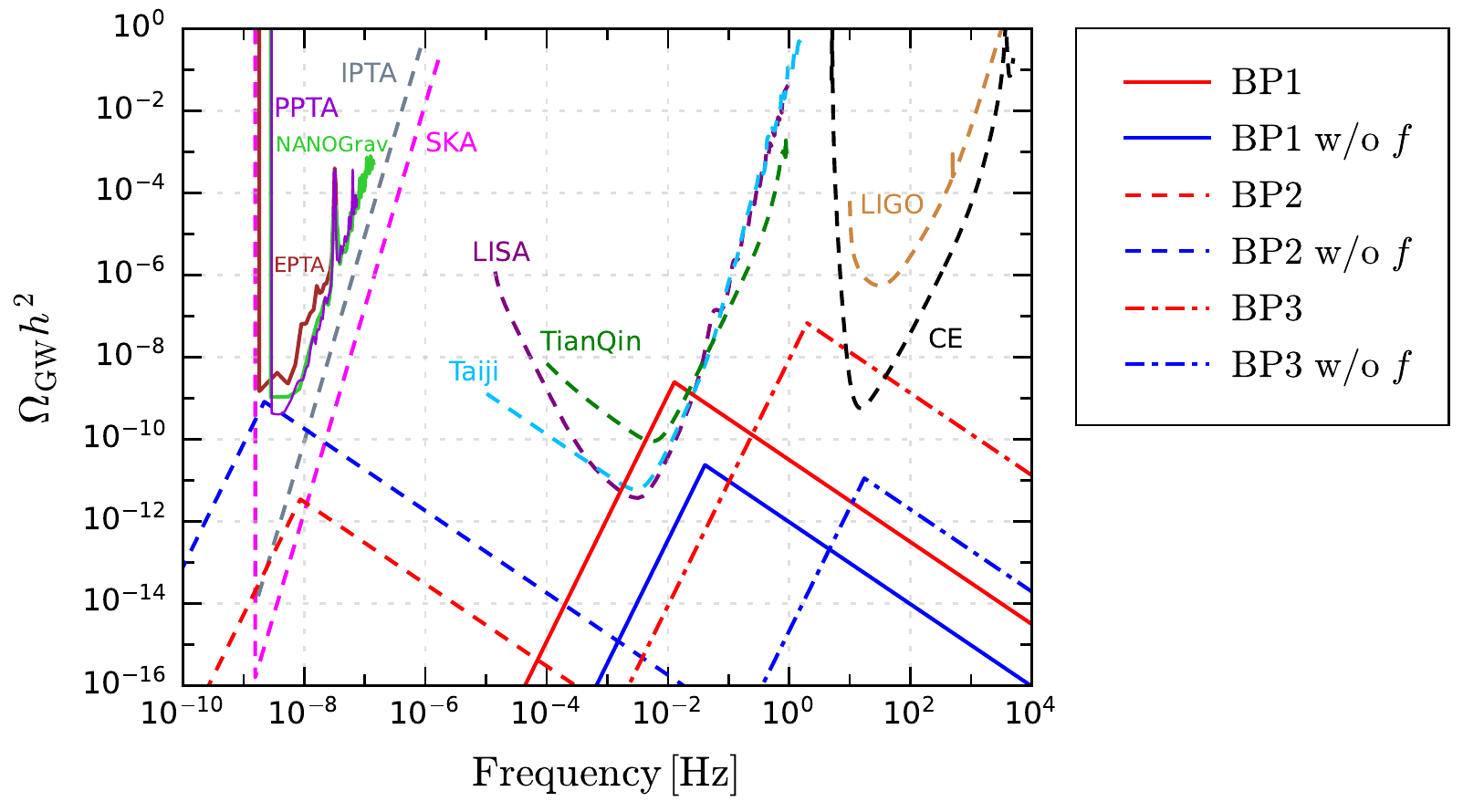}
\caption{SGWB spectra induced by the DWs for the three BPs with (red lines) and without (blue lines) the fermion $f$. Constraints and sensitivity curves of various GW experiments are also plotted.}\label{fig:omega}
\end{figure}

We evaluate the SGWB spectra induced by the DWs for the three BPs with and without the fermion, and the obtained peak frequencies and peak amplitudes are listed in Table~\ref{tab:result}.
The SGWB spectra as functions of GW frequency are demonstrated in Fig.~\ref{fig:omega}, where the constraints and sensitivities curves of current and future GW experiments are also presented.

For BP1 without the fermion, the peak GW frequency is estimated to be $4.08\times 10^{-2}\ \mathrm{Hz}$ with a peak GW amplitude of $\Omega_\mathrm{GW}(f_\mathrm{peak}) h^2 \simeq 2.39\times 10^{-11}$.
Nevertheless, including the fermion in BP1 would decrease both the DW annihilation temperature $T_\mathrm{ann}$ and the peak frequency $f_\mathrm{peak}$ by a factor of $\sim 3.19$, increasing the peak amplitude by 2 orders of magnitude.
As a result, future space-borne GW experiments, such as Laser Interferometer Space Antenna (LISA) \cite{LISA:2017pwj}, 
TianQin \cite{Liang:2021bde}, 
and Taiji \cite{Ruan:2018tsw}, have larger probabilities of probing the latter scenario.

Furthermore, BP2 with or without the fermion predicts a peak frequency around nanohertz, which falls within the sensitive band of PTA experiments, including the North American Nanohertz Observatory for Gravitational Waves (NANOGrav) \cite{NANOGRAV:2018hou},
the European Pulsar Timing Array (EPTA) \cite{Lentati:2015qwp}, 
the Parkes Pulsar Timing Array (PPTA) \cite{Shannon:2015ect}, 
the International Pulsar Timing Array (IPTA) \cite{Hobbs:2009yy}, and
the Square Kilometer Array (SKA) \cite{Janssen:2014dka}.
Compared to BP2 without $f$, BP2 exhibits a reduction in the peak amplitude by 2 orders of magnitude, because including the fermionic contributions would increase both $T_\mathrm{ann}$ and $f_\mathrm{peak}$ by a factor of $\sim 3.93$.
The predicted GW spectra from both cases could be tested by future IPTA and SKA experiments.

Finally, BP3 without the fermion leads to $f_\mathrm{peak} \simeq 17.4\ \mathrm{Hz}$, and the SGWB spectrum lies within the sensitive band of ground-based GW experiments, such as LIGO, Virgo, KAGRA \cite{KAGRA:2013rdx}, and the Cosmic Explorer (CE) \cite{LIGOScientific:2016wof}.
However, the predicted peak amplitude $\Omega_\mathrm{GW}(f_\mathrm{peak}) h^2 \simeq \num{1.14e-11}$ is too low to be probed in these experiments.
Nonetheless, when the effect of the fermion is included, the peak amplitude increases to $\num{6.82e-8}$, which is promising to be detected by the future CE experiment.
Because of $H \propto T^2$ at the radiation-dominated era, a decrease of $T_\mathrm{ann}$ by 1 order of magnitude would reduce $H(T_\mathrm{ann})$ by 2 orders of magnitude and hence increase $\Omega_{\mathrm{GW}}(f_{\mathrm{peak}})\big|_{T_{\mathrm{ann}}}$ by 4 orders of magnitude according to Eq.~\eqref{eq:OmegaGW_peak_Tann}.
This explains the great difference between the GW spectra of BP3 and BP3 w/o $f$.

\begin{figure}[!t]
\centering
\includegraphics[width=0.32\textwidth]{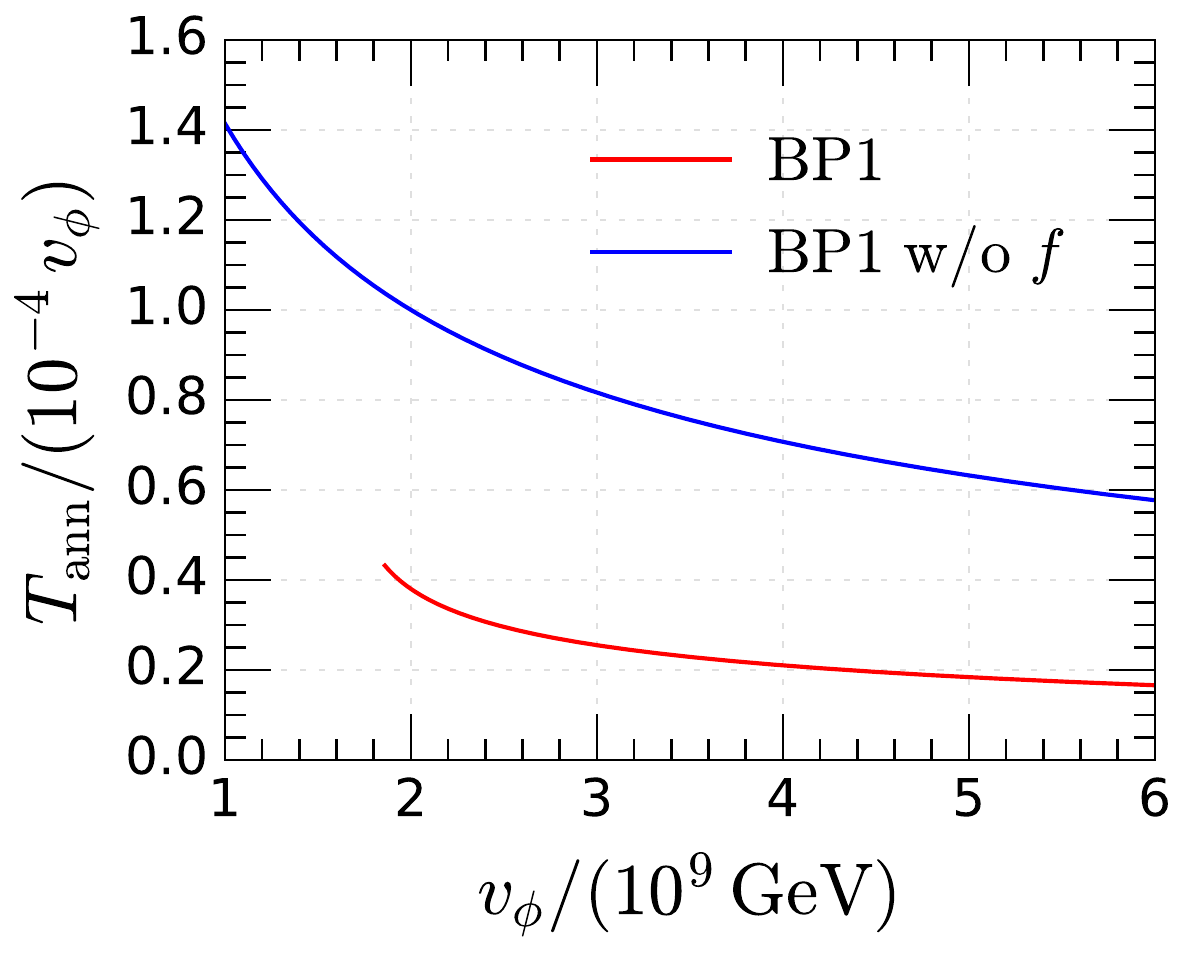}
\includegraphics[width=0.31\textwidth]{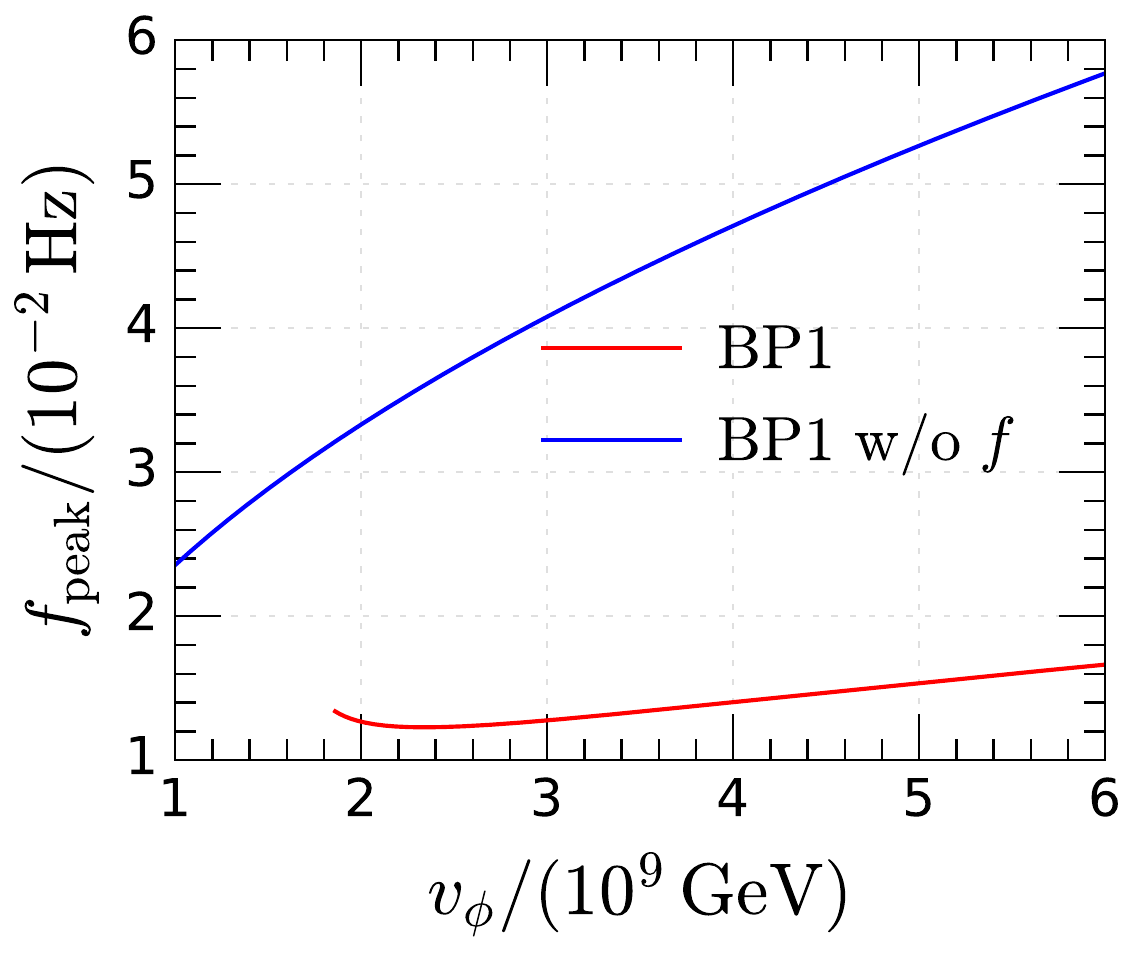}
\includegraphics[width=0.34\textwidth]{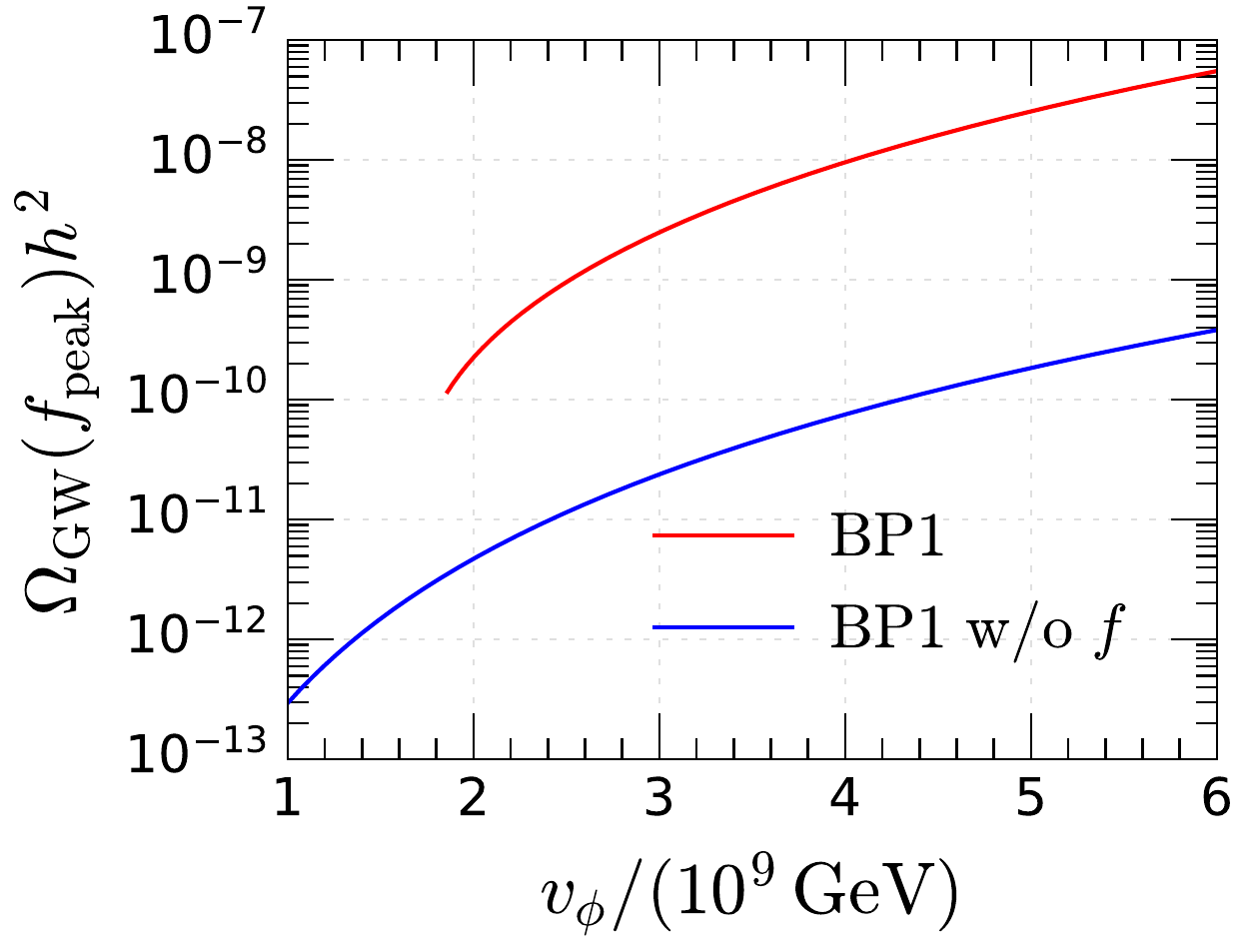}
\includegraphics[width=0.31\textwidth]{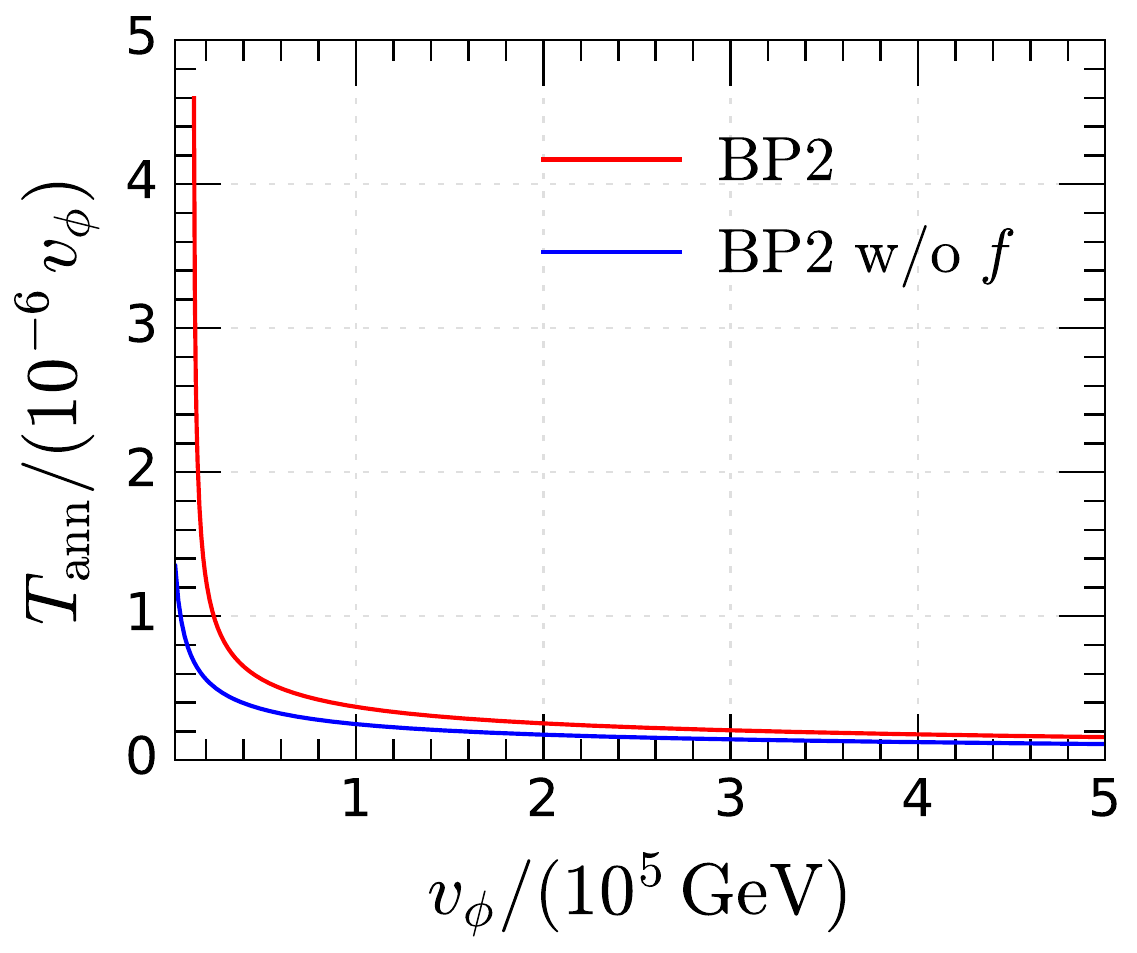}
\includegraphics[width=0.32\textwidth]{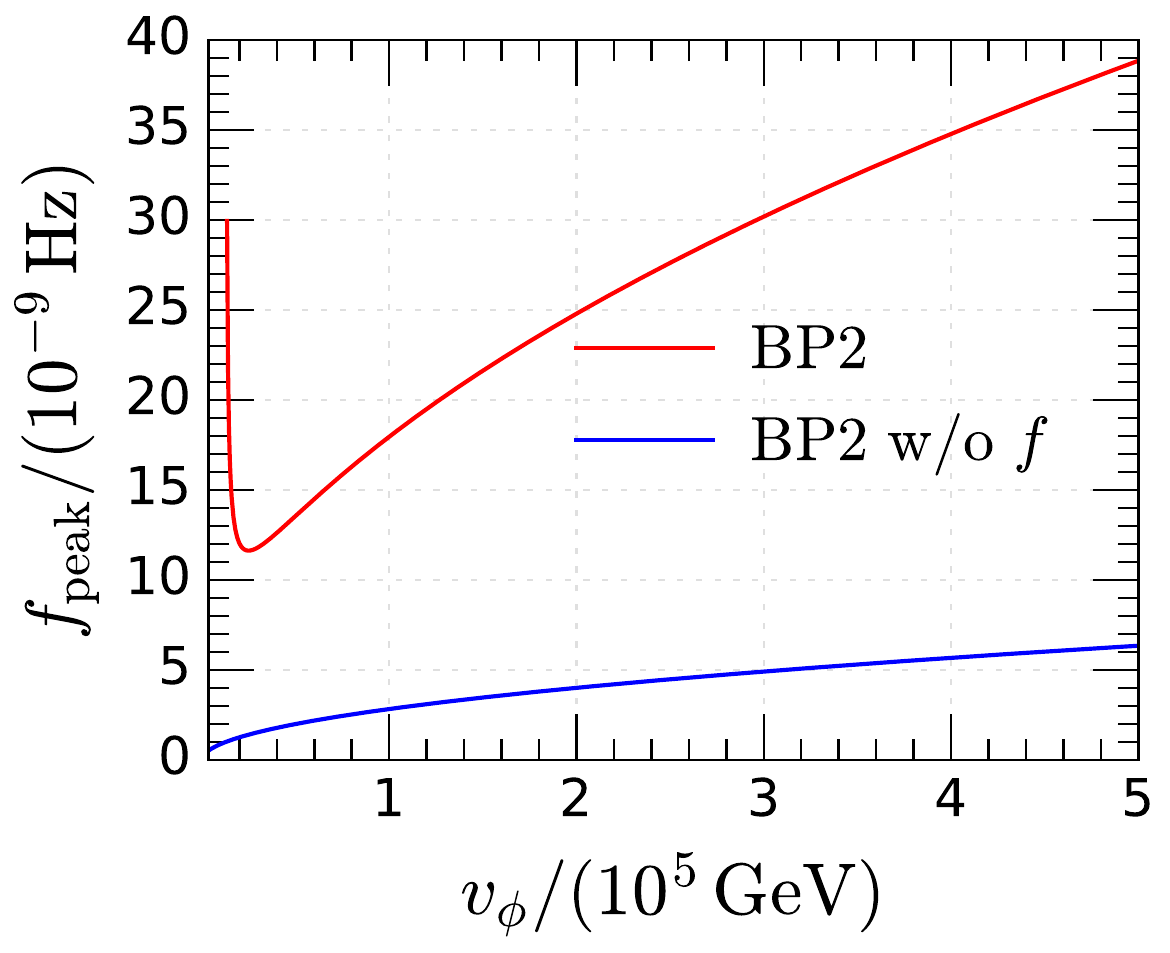}
\includegraphics[width=0.35\textwidth]{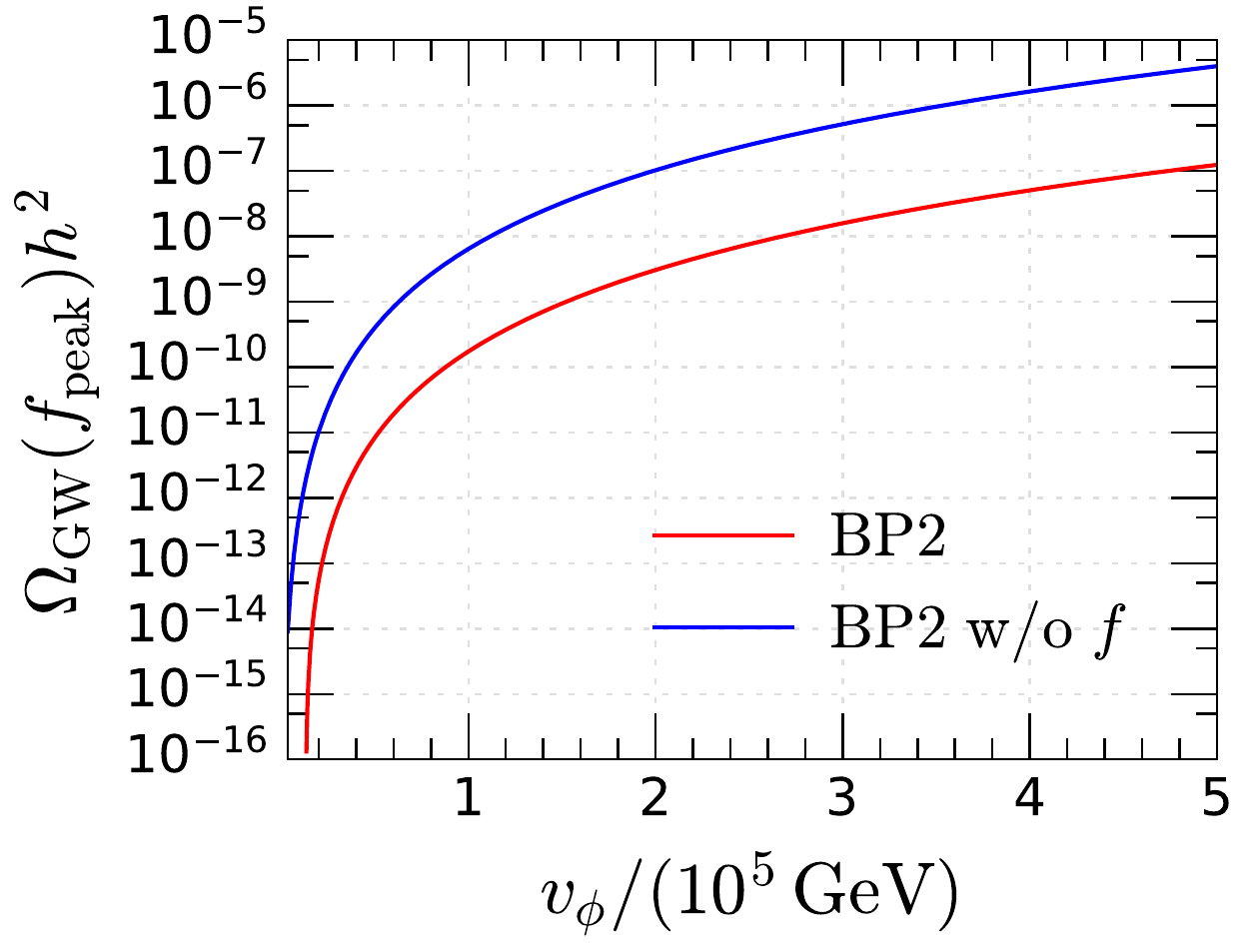}
\includegraphics[width=0.315\textwidth]{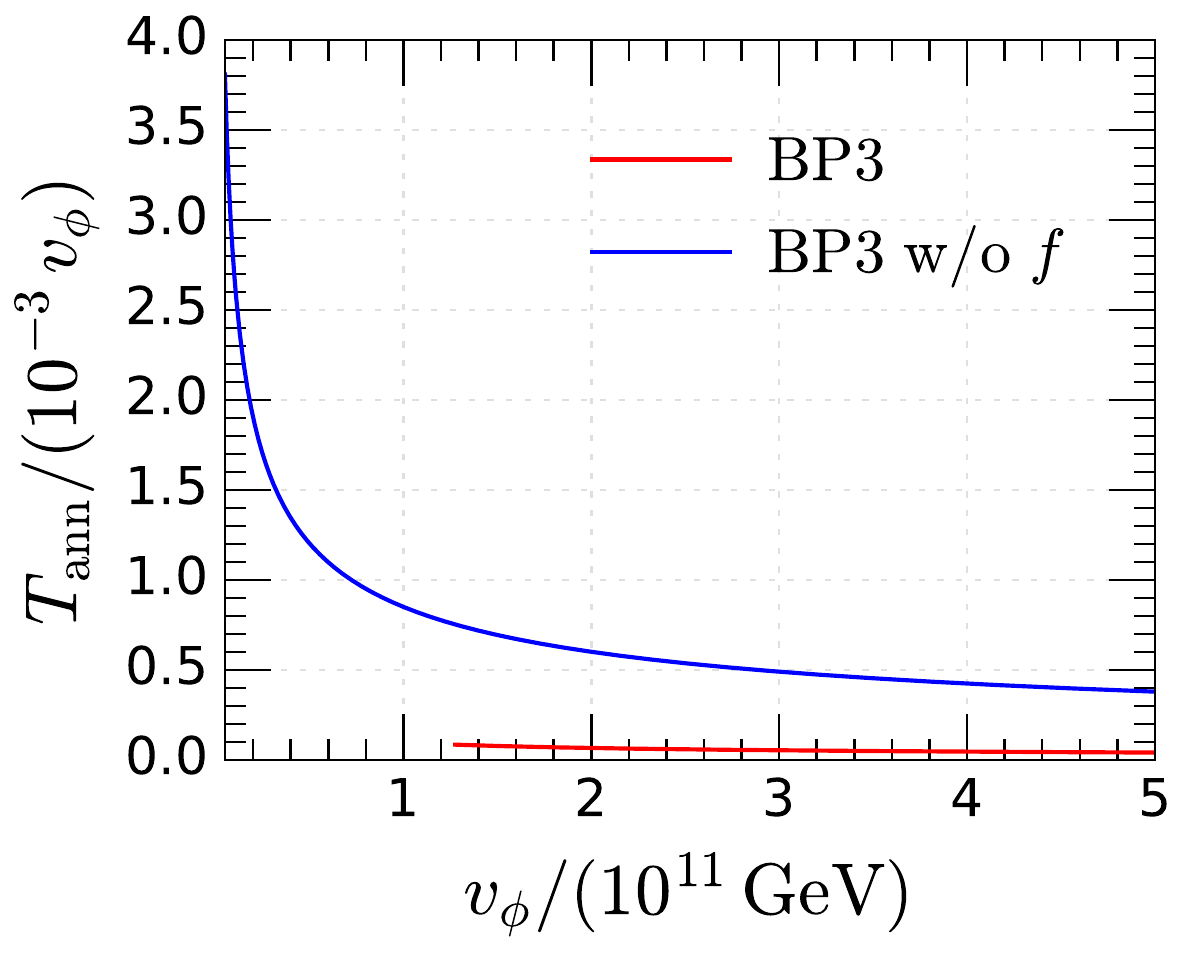}
\includegraphics[width=0.315\textwidth]{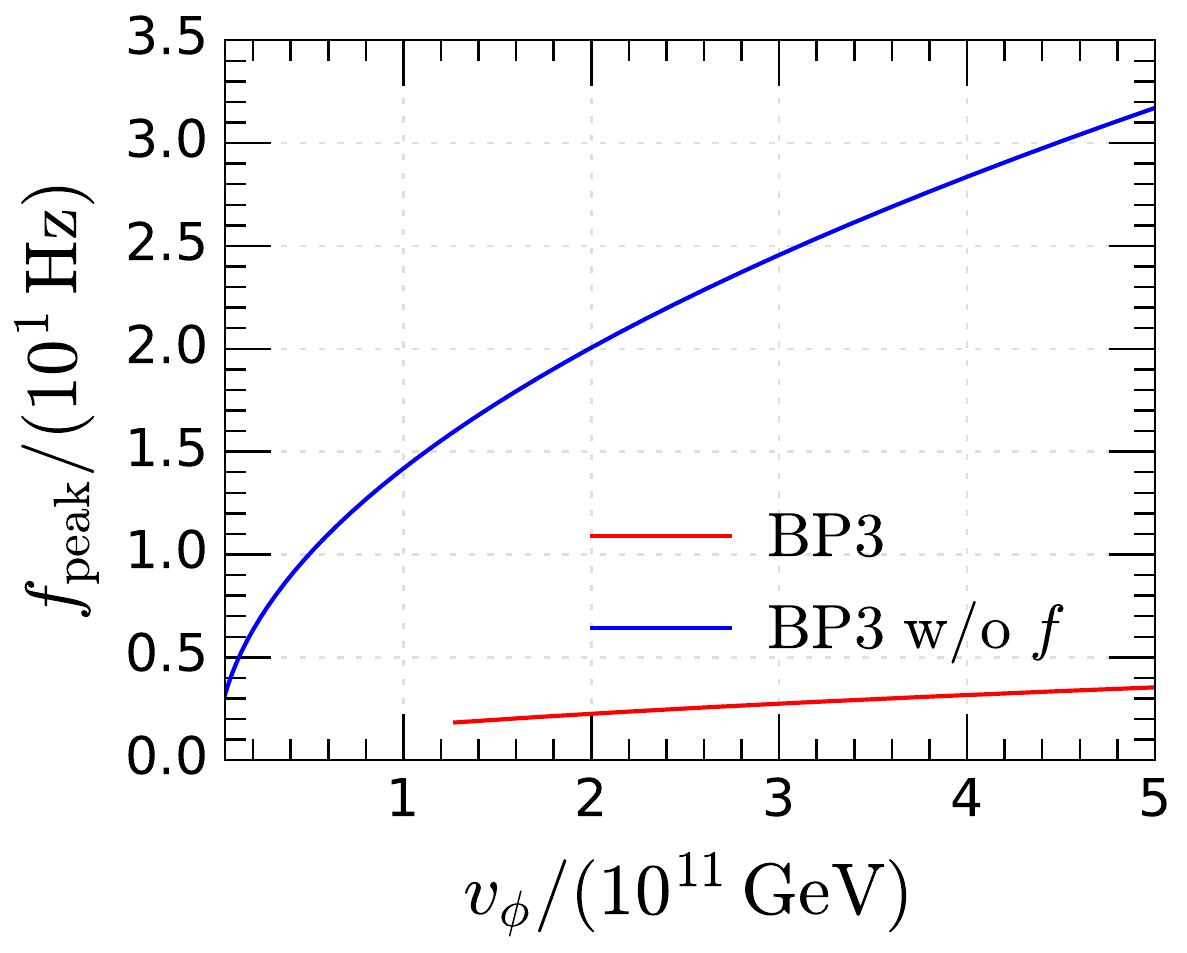}
\includegraphics[width=0.34\textwidth]{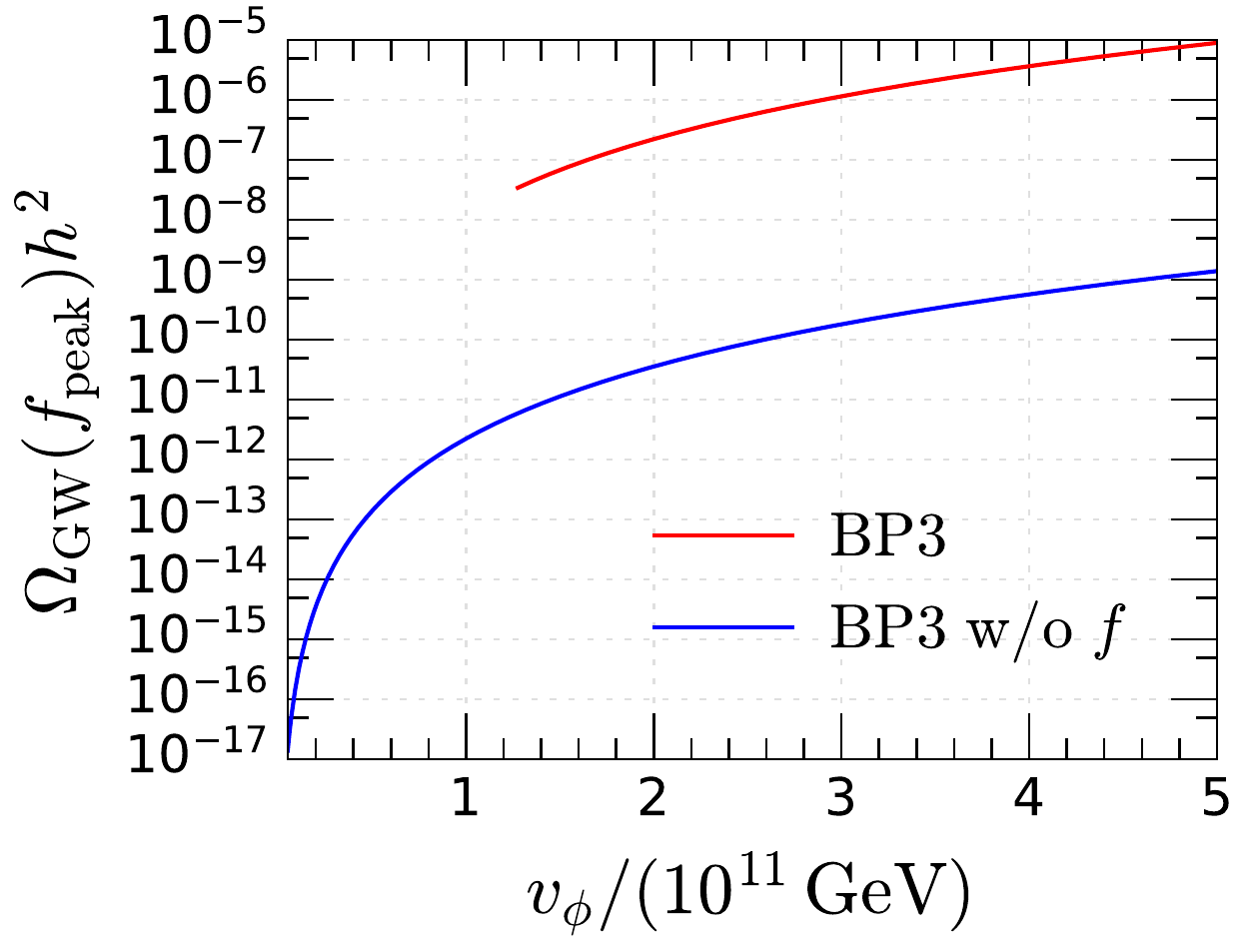}
\caption{$ T_\mathrm{ann}/v_\phi $, $ \Omega_\mathrm{GW}(f_\mathrm{peak})h^2 $, and $ f_\mathrm{peak} $ as functions of $ v_\phi $, with the other parameters fixed as in the three BPs. The left, middle, and right columns correspond to BP1, BP2, and BP3, respectively.
The red and blue curves denote the results for the cases with and without the fermion, respectively.}\label{fig:vphi}
\end{figure}

Below, we investigate how the VEV $v_\phi$ affects the results.
As $ v_\phi $ varies, the curves of $ p_\mathrm{T} $ in Fig.~\ref{fig:ann} would shift compared to those of $ p_\mathrm{V} + F_f $, leading to modifications in the DW annihilation temperature $T_\mathrm{ann}$, which consequently affect both $f_\mathrm{peak}$ and $\Omega_\mathrm{GW}(f_\mathrm{peak}) h^2$.
Based on the BPs listed in Table~\ref{tab:parameter}, we vary the value of $v_\phi$ while keeping the rest parameters $\lambda_3/v_\phi$, $\lambda_\phi$, $y$, and $m_f/v_\phi$ unchanged, and show $T_\mathrm{ann}$, $f_\mathrm{peak}$, and $\Omega_\mathrm{GW}(f_\mathrm{peak}) h^2$ as functions of $v_\phi$ in Fig.~\ref{fig:vphi}.
The red curves, corresponding to the scenario including the fermionic effects, display clear initial points.
This is because if $v_\phi$ is too small, the resulting $p_\mathrm{T}$ in the scaling regime would fall below $p_\mathrm{V}+F_f $ at all temperatures, and the DW annihilation temperature $T_\mathrm{ann}$ would not exist.
In this case, actually, the DWs would collapse rapidly before entering the scaling regime, and their energy density would be insufficient to generate significant GW signals.
On the other hand, the blue curves for the scenarios without the fermion extent to lower values of $v_\phi$, because $p_\mathrm{V}$ is independent of the temperature, ensuring the existence of $T_\mathrm{ann}$.

For large values of $v_\phi$ in the scenario with the fermion, $T_\mathrm{ann}$ is determined at low temperatures where $p_\mathrm{V}$ is basically temperature independent, similar to the scenario without the fermion.
Consequently, the values of $T_\mathrm{ann}/v_\phi$ in both scenarios tend to be constant for large $v_\phi$, and their difference comes from the fermionic contribution to the potential bias in $V_\mathrm{CW}$.
Since $f_\mathrm{peak}$ is positively correlated with $T_\mathrm{ann}$, a constant $T_\mathrm{ann}/v_\phi$ implies that $f_\mathrm{peak}$ is also positively correlated with $v_\phi$, as shown in the middle column of Fig.~\ref{fig:vphi} for large $v_\phi$.
Furthermore, because an increase in $v_\phi$ would enhance the DW tension $\sigma_\mathrm{DW}$, $\Omega_\mathrm{GW}(f_\mathrm{peak}) h^2$ demonstrates a significant and consistent increase with $v_\phi$ in all considered scenarios.

\section{Renormalization scale dependence}\label{sec:muR}

In the previous analyses, we set the parameters and evaluated the effective potential at a fixed renormalization scale $\mu_{\mathrm{R}} = v_\phi$.
%The values of parameters used were also evaluated at $\mu_{\mathrm{R}} = v_\phi$, including $\mu_1^3 \big|_{\mu_\mathrm{R} = v_\phi} = 0$.
For a different $\mu_{\mathrm{R}}$, $V_\mathrm{CW}(\phi)$ is modified due to its explicit dependence on $\mu_{\mathrm{R}}$ and the parameters are shifted according to the renormalization group equations (RGEs).
In this section, we will examine the robustness of our results for the DW annihilation temperature and the resulting SGWB spectrum by evaluating them at another renormalization scale $ \mu_{\mathrm{R}}$.
% By varying $\mu_\mathrm{R}$, we can test the robustness of our results and estimate the theoretical uncertainties.

The all-order zero-temperature effective potential $V_\mathrm{tot}$ is independent of the renormalization scale $\mu_{\mathrm{R}}$, as described by the RGE~\cite{Coleman:1973jx,Andreassen:2014eha,Manohar:2020nzp} that holds order by order,
\begin{equation}
\label{eq:RGE:V_tot}
\left( \frac{\partial}{\partial t}+\sum_{i}\beta_i\frac{\partial}{\partial \lambda_i} -\gamma_\phi\phi\frac{\partial}{\partial \phi}\right) V_{\mathrm{tot}} =0,
\end{equation}
where
\begin{equation}
t \equiv \frac{1}{16\pi^2}\ln\frac{\mu_\mathrm{R}}{\mu_\mathrm{R,0}},
\end{equation}
and the reference scale can be chosen as $\mu_\mathrm{R,0} = v_\phi$.
The parameters in this work are $\lambda_i = \lbrace \mu_1^3,\, \mu_\phi^2,\, \mu_3,\, \lambda_\phi\rbrace$ with $\beta$ functions
\begin{equation}
\beta_i \equiv \frac{\mathrm{d}\lambda_i}{\mathrm{d}t}.
\end{equation}
$ \gamma_\phi $ is the anomalous dimension of the scalar field $\phi$ defined by
\begin{equation}
\frac{\mathrm{d}\phi}{\mathrm{d}t} = -\gamma_\phi\phi.
\end{equation}
Equation~\eqref{eq:RGE:V_tot} dictates that the effect of shifting both the running parameters and the explicit scale dependence of the effective potential according to the RGE can be compensated by a rescaling of the renormalized field $\phi$ according to its anomalous dimension. As a mathematical fact, for a general potential $V(\phi)$, rescaling its variable $\phi$ only shifts the location of the minima in the field space but not the potential value of the minima. Therefore, evaluating the effective potential with parameters and explicit scale dependence set at any scale results in the same potential bias between minima. This ensures the theoretical robustness of our results.

In the following, we demonstrate how such a cancellation of scale dependence works in our calculation. %Since the $\beta$-functions and the anomalous dimension begin 
At the one-loop level, Eq.~\eqref{eq:RGE:V_tot} reduces to 
\begin{equation}
\left( \sum_{i}\beta_i\frac{\partial}{\partial \lambda_i} -\gamma_\phi\phi\frac{\partial}{\partial \phi}\right) V_0 + \frac{\partial V_{\mathrm{CW}}}{\partial t} \simeq 0.
\end{equation}
The one-loop result for the anomalous dimension is $\gamma_\phi = 2y^2$, while
$\beta$ functions that govern the running of the parameters are given by~\cite{Manohar:2020nzp}
\begin{subequations}
\begin{eqnarray}
\frac{\mathrm{d}\mu_1^3}{\mathrm{d}t} &=& -2\mu_\phi^2\mu_3-8ym_f^3+\mu_1^3\gamma_\phi,
\label{eq:mu1_t}
\\
\frac{\mathrm{d}\mu_\phi^2}{\mathrm{d}t} &=& -4\mu_3^2+6\mu_\phi^2\lambda_\phi+24y^2m_f^2+2\mu_\phi^2\gamma_\phi,
\\
\frac{\mathrm{d}\mu_3}{\mathrm{d}t}&=& 18\mu_3\lambda_\phi-24y^3m_f+3\mu_3\gamma_\phi,
\label{eq:mu3_t}
\\
\frac{\mathrm{d}\lambda_\phi}{\mathrm{d}t}&=& 18\lambda_\phi^2-8y^4+4\lambda_\phi\gamma_\phi,
\\
\frac{\mathrm{d}y}{\mathrm{d}t} &=& 5y^3,
\\
\frac{\mathrm{d}m_f}{\mathrm{d}t} &=& 3y^2m_f.
\end{eqnarray}
\end{subequations}
%By solving these RGEs, we can determine the running parameters at any renormalization scale $\mu_\mathrm{R}$.
%For the BPs, the initial values of the parameters at $\mu_{\mathrm{R,0}} = v_\phi$ are provided in Table \ref{tab:parameter} with $\mu_1^3( \mu_{\mathrm{R,0}}) = 0$ and $\mu_\phi^2( \mu_{\mathrm{R,0}})$ determined by Eq.~\eqref{eq:eq-6}.
Note that although we have set the linear term $\mu_1^3( \mu_{\mathrm{R,0}}) = 0$, it becomes nonzero at $\mu_\mathrm{R} \neq \mu_\mathrm{R,0}$, induced by the $\mathbb{Z}_2$-violating couplings $y$ and $\mu_3$.

%First of all, the spectra are determined by $T_{\mathrm{ann}}$, which corresponds to the intersection point of $p_{\mathrm{T}}$ and $p_{\mathrm{V}}\sim V_{\mathrm{bias}}$. 
The physical results in this work mainly depend on the DW tension $\sigma$ and the potential bias $V_{\mathrm{bias}}$. The former is insensitive to $\mu_{\rm R}$ because it is mostly determined by the tree-level potential.
To the leading order, the contributions to the potential bias from various parts of the zero-temperature effective potential can be estimated as
\begin{subequations}
\begin{eqnarray}
\dfrac{\mathrm{d}V_{0,\mathrm{bias}}}{\mathrm{d}t}
%\simeq \left. 
%\dfrac{\mathrm{d}V_0}{\mathrm{d}t}\right|_{\phi=\phi_+}^{\phi=\phi_-}
&\simeq& \left. \left( \dfrac{\mathrm{d} \mu_1^3}{\mathrm{d} t}\phi+\frac{1}{3}\dfrac{\mathrm{d} \mu_3}{\mathrm{d} t}\phi^3 \right) \right|_{\phi=\phi_+}^{\phi=\phi_-}
,
\\
\dfrac{\mathrm{d}V_{\mathrm{CW},\phi,\mathrm{bias}}}{\mathrm{d}t}
&\simeq& \left.-\frac{1}{2} m_\phi^4\right|_{\phi=\phi_+}^{\phi=\phi_-}
,\\
\dfrac{\mathrm{d}V_{\mathrm{CW},f,\mathrm{bias}}}{\mathrm{d}t}
&\simeq& \left. 2\left(m_f+y\phi\right)^4\right|_{\phi=\phi_+}^{\phi=\phi_-},
\end{eqnarray}
\end{subequations}
where $V_{0,\mathrm{bias}}$, $V_{\mathrm{CW},\phi,\mathrm{bias}}$, and $V_{\mathrm{CW},f,\mathrm{bias}}$ are the contributions from the tree potential $V_0$, the scalar field contribution to $V_{\mathrm{CW}}$, and the fermionic contribution to $V_{\mathrm{CW}}$, respectively. 
According to Eqs.~\eqref{eq:mu1_t} and \eqref{eq:mu3_t}, we get
\begin{equation}
\dfrac{\mathrm{d}V_{\mathrm{bias}}}{\mathrm{d}t}\simeq
\dfrac{\mathrm{d}V_{0,\mathrm{bias}}}{\mathrm{d}t}+ \dfrac{\mathrm{d}V_{\mathrm{CW},\phi,\mathrm{bias}}}{\mathrm{d}t} +\dfrac{\mathrm{d}V_{\mathrm{CW},f,\mathrm{bias}}}{\mathrm{d}t}
\simeq 0
.\end{equation}
Therefore, $V_{\mathrm{bias}}$ is almost invariant against a varying $ \mu_\mathrm{R} $, and so are the annihilation temperature $ T_{\mathrm{ann}} $ and the SGWB spectrum.

The above estimation works at the one-loop order. The cancellation of scale dependence may be destroyed if the neglected higher-order terms are important. We assess this issue by running the parameters and evaluating the effective potential numerically at various $\mu_\mathrm{R}$ to investigate the $\mu_\mathrm{R}$ dependence of the DW annihilation temperature $T_\mathrm{ann}$ and the SGWB spectrum.
The relative deviations of $T_\mathrm{ann}$ and the peak GW amplitude $\Omega_\mathrm{GW}(f_\mathrm{peak})$ at a scale $\mu_\mathrm{R}$, relative to $\mu_\mathrm{R,0} = v_\phi$, are defined by
\begin{eqnarray}
\delta_T(\mu_\mathrm{R}) &= &\frac{T_\mathrm{ann} |_{\mu_\mathrm{R}} - T_\mathrm{ann}|_{\mu_\mathrm{R,0}}}{T_\mathrm{ann}|_{\mu_\mathrm{R,0}}},
\\
\delta_\Omega(\mu_\mathrm{R}) &=& \frac{\Omega_\mathrm{GW}(f_\mathrm{peak})|_{\mu_\mathrm{R}}-\Omega_\mathrm{GW}(f_\mathrm{peak})|_{\mu_\mathrm{R,0}}}{\Omega_\mathrm{GW}(f_\mathrm{peak})|_{\mu_\mathrm{R,0}}}.
\end{eqnarray}
In Table~\ref{tab:muR}, we present the relative deviations $\delta_T(\mu_\mathrm{R})$ and $\delta_\Omega(\mu_\mathrm{R})$ calculated at $\mu_\mathrm{R} = v_\phi/2$ and $2v_\phi$ for the BPs with and without the fermion.
According to Eq.~\eqref{eq:f_peak}, the GW peak frequency $f_\mathrm{peak}$ is basically proportional to $T_\mathrm{ann}$, and their relative deviations are nearly identical.
These deviations represent the theoretical uncertainties of the one-loop calculations.
As shown in Table~\ref{tab:muR}, the relative deviation of $T_\mathrm{ann}$ ranges from $\mathcal{O}(10^{-6})$ to $\mathcal{O}(10^{-2})$, while the largest relative deviation of $\Omega_\mathrm{GW}(f_\mathrm{peak})$ is within 20\%.
Therefore, the dependence of the predicted SGWB spectrum on the renormalization scale $\mu_\mathrm{R}$ is well controlled, and our calculations remain sufficiently robust.

\begin{table}[!t]
\caption{Relative deviations $ \delta_T(\mu_\mathrm{R})$ and $\delta_\Omega(\mu_\mathrm{R})$ for the BPs with $\mu_\mathrm{R}=v_\phi/2$ and $2v_\phi$.}\label{tab:muR}
\belowrulesep=0pt
\aboverulesep=0pt
\centering
\renewcommand{\arraystretch}{1.5}
\setlength\tabcolsep{.8em}
\begin{tabular}{ccccc}
\toprule[1.2pt]
& $ \delta_T(\mu_\mathrm{R} = v_\phi/2)$ & 
$ \delta_T(\mu_\mathrm{R} = 2v_\phi)$ & 
$\delta_\Omega(\mu_\mathrm{R} = v_\phi/2)$ &
$\delta_\Omega(\mu_\mathrm{R} = 2v_\phi)$\\
\midrule
BP1& $ 5.40\times 10^{-2} $ & $ -1.11\times 10^{-2} $ & $-1.90\times 10^{-1}$ & $4.61\times 10^{-2}$\\
BP1 w/o $ f $ & $-2.96\times 10^{-3}$ & $-7.54\times 10^{-3}$ & $1.17\times 10^{-2}$ & $3.11\times 10^{-2}$\\
\midrule
BP2 & $ 5.20\times 10^{-5}$ & $ -1.64\times 10^{-4}$ & $-4.19\times 10^{-4}$ & $1.04\times 10^{-3}$\\
BP2 w/o $ f $& $-6.05\times 10^{-6}$ & $-8.84\times 10^{-5}$ & $-1.87\times 10^{-4}$ & $7.40\times 10^{-4}$\\
\midrule
BP3 & $ -9.72\times 10^{-3}$ & $ 2.09\times 10^{-2}$ & $3.96\times 10^{-2}$ & $-7.91\times 10^{-2}$\\
BP3 w/o $ f $ & $-6.10\times 10^{-6}$ & $-8.61\times 10^{-5}$ & $-1.87\times 10^{-4}$ & $ 7.31\times 10^{-4}$\\
\bottomrule[1.2pt]
\end{tabular}
\end{table}

\section{Summary}
\label{sec:summary}

In this paper, we study the evolution of collapsing domain walls affected by the thermal corrections from matter particles and explore the impact on the induced gravitational waves.
We begin by considering a simple model with a real scalar field $ \phi $ coupled to a Dirac spinor field $ f $.
The scalar potential is assumed to respect an approximate $\mathbb{Z}_2$ symmetry, which is slightly violated by a $\phi^3$ term at the tree level and by the Yukawa coupling between $\phi$ and $f$ at the one-loop level.
The $\mathbb{Z}_2$ symmetry results in DWs generated after a second-order phase transition via the Kibble mechanism, while the potential bias $V_\mathrm{bias}$ between the true and false vacua arising from the $\mathbb{Z}_2$-violating terms provides a pressure $p_\mathrm{V} \sim V_\mathrm{bias}$ that drives the collapse of DWs.

In contrast to previous studies that only consider a temperature-independent bare $V_\mathrm{bias}$, we incorporate both quantum and thermal corrections arising from the Yukawa coupling with a fermion field to investigate the temperature dependence of $V_\mathrm{bias}$ and its influences.
After constructing the effective potential with zero-temperature Coleman-Weinberg corrections and finite-temperature corrections at the one-loop level, we systematically compute the temperature dependence of key physical quantities for the evolution of DWs, including the DW tension $\sigma_\mathrm{DW}$, the tension force $p_\mathrm{T}$, the pressure $p_\mathrm{V}$, and the friction force $F_f$. The Coleman-Weinberg potential contributed by the Yukawa coupling shifts the temperature-independent part of $V_\mathrm{bias}$.
On the other hand, because of the thermal corrections from the fermion, the pressure $p_\mathrm{V}$ is proportional to $T^2$ at sufficiently high temperatures, leading to further differences in the determined DW annihilation temperature $T_\mathrm{ann}$ between the scenarios with and without the fermionic contribution.

The frequency spectra of the SGWB generated by collapsing DWs in three BPs are further estimated, and we find that the differences in $T_\mathrm{ann}$ could lead to substantial modifications in the GW spectra.
For the selected parameter sets, the GW spectra corresponding to BP1, BP2, and BP3 lie within the observational windows of space-borne interferometers, PTAs, and ground-based interferometers, respectively.
Therefore, these characteristic spectral modifications could potentially be verified by future GW experiments.
Remarkably, the DW annihilation temperature for BP3 decreases by a factor of $10$ compared to the case without the fermionic contributions, leading to a dramatic enhancement in the peak amplitude of the GW spectrum by 4 orders of magnitude, which significantly improves the detectability prospects in future GW experiments.
Moreover, we investigate the dependence of $T_\mathrm{ann}$, $f_\mathrm{peak}$, and $\Omega_\mathrm{GW}(f_\mathrm{peak})h^2$ on the  VEV $v_\phi$, revealing that increasing $v_\phi$ would significantly enhance the amplitude of the GW spectrum. 

Finally, we investigate the dependence of SGWB spectra on the renormalization scale $ \mu_\mathrm{R}$.
The running parameters are obtained by solving the related RGEs, and the effective potential is calculated for different values of $ \mu_\mathrm{R}$.
By varying $\mu_\mathrm{R}$ to $v_\phi/2$ and $v_\phi$, we find that the relative deviations of the DW annihilation temperature and the peak GW frequency for the BPs are within $6\%$ and $20\%$, respectively.
This implies that the theoretical uncertainties of our one-loop calculations are manageable, and the obtained results are reliable.

\begin{acknowledgments}

This work is supported by the Guangzhou Science and Technology Planning Project under Grant No.~2024A04J4026.

\end{acknowledgments}

\bibliographystyle{utphys}
\bibliography{ref}

\providecommand{\href}[2]{#2}\begingroup\raggedright\begin{thebibliography}{10}

\bibitem{LIGOScientific:2016aoc}
{\bfseries LIGO Scientific, Virgo} Collaboration, B.~P. Abbott {\em et~al.},
  ``{Observation of Gravitational Waves from a Binary Black Hole Merger},''
  \href{http://dx.doi.org/10.1103/PhysRevLett.116.061102}{{\em Phys. Rev.
  Lett.} {\bfseries 116} (2016) 061102},
  \href{http://arxiv.org/abs/1602.03837}{{\ttfamily arXiv:1602.03837 [gr-qc]}}.

\bibitem{LIGOScientific:2017vwq}
{\bfseries LIGO Scientific, Virgo} Collaboration, B.~P. Abbott {\em et~al.},
  ``{GW170817: Observation of Gravitational Waves from a Binary Neutron Star
  Inspiral},'' \href{http://dx.doi.org/10.1103/PhysRevLett.119.161101}{{\em
  Phys. Rev. Lett.} {\bfseries 119} (2017) 161101},
  \href{http://arxiv.org/abs/1710.05832}{{\ttfamily arXiv:1710.05832 [gr-qc]}}.

\bibitem{Caprini:2018mtu}
C.~Caprini and D.~G. Figueroa, ``{Cosmological Backgrounds of Gravitational
  Waves},'' \href{http://dx.doi.org/10.1088/1361-6382/aac608}{{\em Class.
  Quant. Grav.} {\bfseries 35} (2018) 163001},
  \href{http://arxiv.org/abs/1801.04268}{{\ttfamily arXiv:1801.04268
  [astro-ph.CO]}}.

\bibitem{Renzini:2022alw}
A.~I. Renzini, B.~Goncharov, A.~C. Jenkins, and P.~M. Meyers, ``{Stochastic
  Gravitational-Wave Backgrounds: Current Detection Efforts and Future
  Prospects},'' \href{http://dx.doi.org/10.3390/galaxies10010034}{{\em
  Galaxies} {\bfseries 10} (2022) 34},
  \href{http://arxiv.org/abs/2202.00178}{{\ttfamily arXiv:2202.00178 [gr-qc]}}.

\bibitem{KAGRA:2013rdx}
{\bfseries KAGRA, LIGO Scientific, Virgo, VIRGO} Collaboration, B.~P. Abbott
  {\em et~al.}, ``{Prospects for observing and localizing gravitational-wave
  transients with Advanced LIGO, Advanced Virgo and KAGRA},''
  \href{http://dx.doi.org/10.1007/s41114-020-00026-9}{{\em Living Rev. Rel.}
  {\bfseries 21} (2018) 3}, \href{http://arxiv.org/abs/1304.0670}{{\ttfamily
  arXiv:1304.0670 [gr-qc]}}.

\bibitem{LIGOScientific:2016wof}
{\bfseries LIGO Scientific} Collaboration, B.~P. Abbott {\em et~al.},
  ``{Exploring the Sensitivity of Next Generation Gravitational Wave
  Detectors},'' \href{http://dx.doi.org/10.1088/1361-6382/aa51f4}{{\em Class.
  Quant. Grav.} {\bfseries 34} (2017) 044001},
  \href{http://arxiv.org/abs/1607.08697}{{\ttfamily arXiv:1607.08697
  [astro-ph.IM]}}.

\bibitem{NANOGRAV:2018hou}
{\bfseries NANOGRAV} Collaboration, Z.~Arzoumanian {\em et~al.}, ``{The
  NANOGrav 11-year Data Set: Pulsar-timing Constraints On The Stochastic
  Gravitational-wave Background},''
  \href{http://dx.doi.org/10.3847/1538-4357/aabd3b}{{\em Astrophys. J.}
  {\bfseries 859} (2018) 47}, \href{http://arxiv.org/abs/1801.02617}{{\ttfamily
  arXiv:1801.02617 [astro-ph.HE]}}.

\bibitem{Lentati:2015qwp}
L.~Lentati {\em et~al.}, ``{European Pulsar Timing Array Limits On An Isotropic
  Stochastic Gravitational-Wave Background},''
  \href{http://dx.doi.org/10.1093/mnras/stv1538}{{\em Mon. Not. Roy. Astron.
  Soc.} {\bfseries 453} (2015) 2576--2598},
  \href{http://arxiv.org/abs/1504.03692}{{\ttfamily arXiv:1504.03692
  [astro-ph.CO]}}.

\bibitem{Shannon:2015ect}
R.~M. Shannon {\em et~al.}, ``{Gravitational waves from binary supermassive
  black holes missing in pulsar observations},''
  \href{http://dx.doi.org/10.1126/science.aab1910}{{\em Science} {\bfseries
  349} (2015) 1522--1525}, \href{http://arxiv.org/abs/1509.07320}{{\ttfamily
  arXiv:1509.07320 [astro-ph.CO]}}.

\bibitem{Hobbs:2009yy}
G.~Hobbs {\em et~al.}, ``{The international pulsar timing array project: using
  pulsars as a gravitational wave detector},''
  \href{http://dx.doi.org/10.1088/0264-9381/27/8/084013}{{\em Class. Quant.
  Grav.} {\bfseries 27} (2010) 084013},
  \href{http://arxiv.org/abs/0911.5206}{{\ttfamily arXiv:0911.5206
  [astro-ph.SR]}}.

\bibitem{Janssen:2014dka}
G.~Janssen {\em et~al.}, ``{Gravitational wave astronomy with the SKA},''
  \href{http://dx.doi.org/10.22323/1.215.0037}{{\em PoS} {\bfseries AASKA14}
  (2015) 037}, \href{http://arxiv.org/abs/1501.00127}{{\ttfamily
  arXiv:1501.00127 [astro-ph.IM]}}.

\bibitem{LISA:2017pwj}
{\bfseries LISA} Collaboration, P.~Amaro-Seoane {\em et~al.}, ``{Laser
  Interferometer Space Antenna},''
  \href{http://arxiv.org/abs/1702.00786}{{\ttfamily arXiv:1702.00786
  [astro-ph.IM]}}.

\bibitem{Ruan:2018tsw}
W.-H. Ruan, Z.-K. Guo, R.-G. Cai, and Y.-Z. Zhang, ``{Taiji program:
  Gravitational-wave sources},''
  \href{http://dx.doi.org/10.1142/S0217751X2050075X}{{\em Int. J. Mod. Phys. A}
  {\bfseries 35} (2020) 2050075},
  \href{http://arxiv.org/abs/1807.09495}{{\ttfamily arXiv:1807.09495 [gr-qc]}}.

\bibitem{Liang:2021bde}
Z.-C. Liang, Y.-M. Hu, Y.~Jiang, J.~Cheng, J.-d. Zhang, and J.~Mei, ``{Science
  with the TianQin Observatory: Preliminary results on stochastic
  gravitational-wave background},''
  \href{http://dx.doi.org/10.1103/PhysRevD.105.022001}{{\em Phys. Rev. D}
  {\bfseries 105} (2022) 022001},
  \href{http://arxiv.org/abs/2107.08643}{{\ttfamily arXiv:2107.08643
  [astro-ph.CO]}}.

\bibitem{NANOGrav:2023gor}
{\bfseries NANOGrav} Collaboration, G.~Agazie {\em et~al.}, ``{The NANOGrav 15
  yr Data Set: Evidence for a Gravitational-wave Background},''
  \href{http://dx.doi.org/10.3847/2041-8213/acdac6}{{\em Astrophys. J. Lett.}
  {\bfseries 951} (2023) L8}, \href{http://arxiv.org/abs/2306.16213}{{\ttfamily
  arXiv:2306.16213 [astro-ph.HE]}}.

\bibitem{EPTA:2023fyk}
{\bfseries EPTA, InPTA:} Collaboration, J.~Antoniadis {\em et~al.}, ``{The
  second data release from the European Pulsar Timing Array - III. Search for
  gravitational wave signals},''
  \href{http://dx.doi.org/10.1051/0004-6361/202346844}{{\em Astron. Astrophys.}
  {\bfseries 678} (2023) A50},
  \href{http://arxiv.org/abs/2306.16214}{{\ttfamily arXiv:2306.16214
  [astro-ph.HE]}}.

\bibitem{Reardon:2023gzh}
D.~J. Reardon {\em et~al.}, ``{Search for an Isotropic Gravitational-wave
  Background with the Parkes Pulsar Timing Array},''
  \href{http://dx.doi.org/10.3847/2041-8213/acdd02}{{\em Astrophys. J. Lett.}
  {\bfseries 951} (2023) L6}, \href{http://arxiv.org/abs/2306.16215}{{\ttfamily
  arXiv:2306.16215 [astro-ph.HE]}}.

\bibitem{Xu:2023wog}
H.~Xu {\em et~al.}, ``{Searching for the Nano-Hertz Stochastic Gravitational
  Wave Background with the Chinese Pulsar Timing Array Data Release I},''
  \href{http://dx.doi.org/10.1088/1674-4527/acdfa5}{{\em Res. Astron.
  Astrophys.} {\bfseries 23} (2023) 075024},
  \href{http://arxiv.org/abs/2306.16216}{{\ttfamily arXiv:2306.16216
  [astro-ph.HE]}}.

\bibitem{NANOGrav:2023hvm}
{\bfseries NANOGrav} Collaboration, A.~Afzal {\em et~al.}, ``{The NANOGrav 15
  yr Data Set: Search for Signals from New Physics},''
  \href{http://dx.doi.org/10.3847/2041-8213/acdc91}{{\em Astrophys. J. Lett.}
  {\bfseries 951} (2023) L11},
  \href{http://arxiv.org/abs/2306.16219}{{\ttfamily arXiv:2306.16219
  [astro-ph.HE]}}. [Erratum: Astrophys.J.Lett. 971, L27 (2024), Erratum:
  Astrophys.J. 971, L27 (2024)].

\bibitem{EPTA:2023xxk}
{\bfseries EPTA, InPTA} Collaboration, J.~Antoniadis {\em et~al.}, ``{The
  second data release from the European Pulsar Timing Array - IV. Implications
  for massive black holes, dark matter, and the early Universe},''
  \href{http://dx.doi.org/10.1051/0004-6361/202347433}{{\em Astron. Astrophys.}
  {\bfseries 685} (2024) A94},
  \href{http://arxiv.org/abs/2306.16227}{{\ttfamily arXiv:2306.16227
  [astro-ph.CO]}}.

\bibitem{Kibble:1976sj}
T.~W.~B. Kibble, ``{Topology of Cosmic Domains and Strings},''
  \href{http://dx.doi.org/10.1088/0305-4470/9/8/029}{{\em J. Phys. A}
  {\bfseries 9} (1976) 1387--1398}.

\bibitem{Vilenkin:1981zs}
A.~Vilenkin, ``{Gravitational Field of Vacuum Domain Walls and Strings},''
  \href{http://dx.doi.org/10.1103/PhysRevD.23.852}{{\em Phys. Rev. D}
  {\bfseries 23} (1981) 852--857}.

\bibitem{Vachaspati:2000cq}
T.~Vachaspati, ``{Lectures on cosmic topological defects},'' {\em ICTP Lect.
  Notes Ser.} {\bfseries 4} (2001) 165--202,
  \href{http://arxiv.org/abs/hep-ph/0101270}{{\ttfamily arXiv:hep-ph/0101270}}.

\bibitem{Vilenkin:2000jqa}
A.~Vilenkin and E.~P.~S. Shellard, {\em {Cosmic Strings and Other Topological
  Defects}}.
\newblock Cambridge University Press, 7, 2000.

\bibitem{Zeldovich:1974uw}
Y.~B. Zeldovich, I.~Y. Kobzarev, and L.~B. Okun, ``{Cosmological Consequences
  of the Spontaneous Breakdown of Discrete Symmetry},'' {\em Zh. Eksp. Teor.
  Fiz.} {\bfseries 67} (1974) 3--11.

\bibitem{Press:1989yh}
W.~H. Press, B.~S. Ryden, and D.~N. Spergel, ``{Dynamical Evolution of Domain
  Walls in an Expanding Universe},''
  \href{http://dx.doi.org/10.1086/168151}{{\em Astrophys. J.} {\bfseries 347}
  (1989) 590--604}.

\bibitem{Abbott:1989jw}
L.~F. Abbott and M.~B. Wise, ``{Wormholes and Global Symmetries},''
  \href{http://dx.doi.org/10.1016/0550-3213(89)90503-8}{{\em Nucl. Phys. B}
  {\bfseries 325} (1989) 687--704}.

\bibitem{Coleman:1989zu}
S.~R. Coleman and K.-M. Lee, ``{Wormholes made without massless matter
  fields},'' \href{http://dx.doi.org/10.1016/0550-3213(90)90149-8}{{\em Nucl.
  Phys. B} {\bfseries 329} (1990) 387--409}.

\bibitem{Hiramatsu:2010yz}
T.~Hiramatsu, M.~Kawasaki, and K.~Saikawa, ``{Gravitational Waves from
  Collapsing Domain Walls},''
  \href{http://dx.doi.org/10.1088/1475-7516/2010/05/032}{{\em JCAP} {\bfseries
  05} (2010) 032}, \href{http://arxiv.org/abs/1002.1555}{{\ttfamily
  arXiv:1002.1555 [astro-ph.CO]}}.

\bibitem{Kawasaki:2011vv}
M.~Kawasaki and K.~Saikawa, ``{Study of gravitational radiation from cosmic
  domain walls},'' \href{http://dx.doi.org/10.1088/1475-7516/2011/09/008}{{\em
  JCAP} {\bfseries 09} (2011) 008},
  \href{http://arxiv.org/abs/1102.5628}{{\ttfamily arXiv:1102.5628
  [astro-ph.CO]}}.

\bibitem{Gelmini:1988sf}
G.~B. Gelmini, M.~Gleiser, and E.~W. Kolb, ``{Cosmology of Biased Discrete
  Symmetry Breaking},'' \href{http://dx.doi.org/10.1103/PhysRevD.39.1558}{{\em
  Phys. Rev. D} {\bfseries 39} (1989) 1558}.

\bibitem{Kadota:2015dza}
K.~Kadota, M.~Kawasaki, and K.~Saikawa, ``{Gravitational waves from domain
  walls in the next-to-minimal supersymmetric standard model},''
  \href{http://dx.doi.org/10.1088/1475-7516/2015/10/041}{{\em JCAP} {\bfseries
  10} (2015) 041}, \href{http://arxiv.org/abs/1503.06998}{{\ttfamily
  arXiv:1503.06998 [hep-ph]}}.

\bibitem{Borah:2022wdy}
D.~Borah and A.~Dasgupta, ``{Probing left-right symmetry via gravitational
  waves from domain walls},''
  \href{http://dx.doi.org/10.1103/PhysRevD.106.035016}{{\em Phys. Rev. D}
  {\bfseries 106} (2022) 035016},
  \href{http://arxiv.org/abs/2205.12220}{{\ttfamily arXiv:2205.12220
  [hep-ph]}}.

\bibitem{Coulson:1995nv}
D.~Coulson, Z.~Lalak, and B.~A. Ovrut, ``{Biased domain walls},''
  \href{http://dx.doi.org/10.1103/PhysRevD.53.4237}{{\em Phys. Rev. D}
  {\bfseries 53} (1996) 4237--4246}.

\bibitem{Gleiser:1998na}
M.~Gleiser and R.~Roberts, ``{Gravitational waves from collapsing vacuum
  domains},'' \href{http://dx.doi.org/10.1103/PhysRevLett.81.5497}{{\em Phys.
  Rev. Lett.} {\bfseries 81} (1998) 5497--5500},
  \href{http://arxiv.org/abs/astro-ph/9807260}{{\ttfamily
  arXiv:astro-ph/9807260}}.

\bibitem{Hiramatsu:2013qaa}
T.~Hiramatsu, M.~Kawasaki, and K.~Saikawa, ``{On the estimation of
  gravitational wave spectrum from cosmic domain walls},''
  \href{http://dx.doi.org/10.1088/1475-7516/2014/02/031}{{\em JCAP} {\bfseries
  02} (2014) 031}, \href{http://arxiv.org/abs/1309.5001}{{\ttfamily
  arXiv:1309.5001 [astro-ph.CO]}}.

\bibitem{Nakayama:2016gxi}
K.~Nakayama, F.~Takahashi, and N.~Yokozaki, ``{Gravitational waves from domain
  walls and their implications},''
  \href{http://dx.doi.org/10.1016/j.physletb.2017.05.010}{{\em Phys. Lett. B}
  {\bfseries 770} (2017) 500--506},
  \href{http://arxiv.org/abs/1612.08327}{{\ttfamily arXiv:1612.08327
  [hep-ph]}}.

\bibitem{Saikawa:2017hiv}
K.~Saikawa, ``{A review of gravitational waves from cosmic domain walls},''
  \href{http://dx.doi.org/10.3390/universe3020040}{{\em Universe} {\bfseries 3}
  (2017) 40}, \href{http://arxiv.org/abs/1703.02576}{{\ttfamily
  arXiv:1703.02576 [hep-ph]}}.

\bibitem{Gelmini:2020bqg}
G.~B. Gelmini, S.~Pascoli, E.~Vitagliano, and Y.-L. Zhou, ``{Gravitational wave
  signatures from discrete flavor symmetries},''
  \href{http://dx.doi.org/10.1088/1475-7516/2021/02/032}{{\em JCAP} {\bfseries
  02} (2021) 032}, \href{http://arxiv.org/abs/2009.01903}{{\ttfamily
  arXiv:2009.01903 [hep-ph]}}.

\bibitem{Zhang:2023nrs}
Z.~Zhang, C.~Cai, Y.-H. Su, S.~Wang, Z.-H. Yu, and H.-H. Zhang, ``{Nano-Hertz
  gravitational waves from collapsing domain walls associated with freeze-in
  dark matter in light of pulsar timing array observations},''
  \href{http://dx.doi.org/10.1103/PhysRevD.108.095037}{{\em Phys. Rev. D}
  {\bfseries 108} (2023) 095037},
  \href{http://arxiv.org/abs/2307.11495}{{\ttfamily arXiv:2307.11495
  [hep-ph]}}.

\bibitem{Stojkovic:2005zh}
D.~Stojkovic, K.~Freese, and G.~D. Starkman, ``{Holes in the walls: Primordial
  black holes as a solution to the cosmological domain wall problem},''
  \href{http://dx.doi.org/10.1103/PhysRevD.72.045012}{{\em Phys. Rev. D}
  {\bfseries 72} (2005) 045012},
  \href{http://arxiv.org/abs/hep-ph/0505026}{{\ttfamily arXiv:hep-ph/0505026}}.

\bibitem{Dunsky:2021tih}
D.~I. Dunsky, A.~Ghoshal, H.~Murayama, Y.~Sakakihara, and G.~White, ``{GUTs,
  hybrid topological defects, and gravitational waves},''
  \href{http://dx.doi.org/10.1103/PhysRevD.106.075030}{{\em Phys. Rev. D}
  {\bfseries 106} (2022) 075030},
  \href{http://arxiv.org/abs/2111.08750}{{\ttfamily arXiv:2111.08750
  [hep-ph]}}.

\bibitem{Lazarides:1982tw}
G.~Lazarides and Q.~Shafi, ``{Axion Models with No Domain Wall Problem},''
  \href{http://dx.doi.org/10.1016/0370-2693(82)90506-8}{{\em Phys. Lett. B}
  {\bfseries 115} (1982) 21--25}.

\bibitem{Coleman:1973jx}
S.~R. Coleman and E.~J. Weinberg, ``{Radiative Corrections as the Origin of
  Spontaneous Symmetry Breaking},''
  \href{http://dx.doi.org/10.1103/PhysRevD.7.1888}{{\em Phys. Rev. D}
  {\bfseries 7} (1973) 1888--1910}.

\bibitem{Delaunay:2007wb}
C.~Delaunay, C.~Grojean, and J.~D. Wells, ``{Dynamics of Non-renormalizable
  Electroweak Symmetry Breaking},''
  \href{http://dx.doi.org/10.1088/1126-6708/2008/04/029}{{\em JHEP} {\bfseries
  04} (2008) 029}, \href{http://arxiv.org/abs/0711.2511}{{\ttfamily
  arXiv:0711.2511 [hep-ph]}}.

\bibitem{Quiros:1999jp}
M.~Quiros, ``{Finite temperature field theory and phase transitions},'' in {\em
  {ICTP Summer School in High-Energy Physics and Cosmology}}, pp.~187--259.
\newblock 1, 1999, \href{http://arxiv.org/abs/hep-ph/9901312}{{\ttfamily
  arXiv:hep-ph/9901312}}.

\bibitem{Andreassen:2014eha}
A.~Andreassen, W.~Frost, and M.~D. Schwartz, ``{Consistent Use of Effective
  Potentials},'' \href{http://dx.doi.org/10.1103/PhysRevD.91.016009}{{\em Phys.
  Rev. D} {\bfseries 91} (2015) 016009},
  \href{http://arxiv.org/abs/1408.0287}{{\ttfamily arXiv:1408.0287 [hep-ph]}}.

\bibitem{Stauffer:1978kr}
D.~Stauffer, ``{Scaling theory of percolation clusters},''
  \href{http://dx.doi.org/10.1016/0370-1573(79)90060-7}{{\em Phys. Rept.}
  {\bfseries 54} (1979) 1--74}.

\bibitem{Hindmarsh:1996xv}
M.~Hindmarsh, ``{Analytic scaling solutions for cosmic domain walls},''
  \href{http://dx.doi.org/10.1103/PhysRevLett.77.4495}{{\em Phys. Rev. Lett.}
  {\bfseries 77} (1996) 4495--4498},
  \href{http://arxiv.org/abs/hep-ph/9605332}{{\ttfamily arXiv:hep-ph/9605332}}.

\bibitem{Hindmarsh:2002bq}
M.~Hindmarsh, ``{Level set method for the evolution of defect and brane
  networks},'' \href{http://dx.doi.org/10.1103/PhysRevD.68.043510}{{\em Phys.
  Rev. D} {\bfseries 68} (2003) 043510},
  \href{http://arxiv.org/abs/hep-ph/0207267}{{\ttfamily arXiv:hep-ph/0207267}}.

\bibitem{Garagounis:2002kt}
T.~Garagounis and M.~Hindmarsh, ``{Scaling in numerical simulations of domain
  walls},'' \href{http://dx.doi.org/10.1103/PhysRevD.68.103506}{{\em Phys. Rev.
  D} {\bfseries 68} (2003) 103506},
  \href{http://arxiv.org/abs/hep-ph/0212359}{{\ttfamily arXiv:hep-ph/0212359}}.

\bibitem{Kolb:1990vq}
E.~W. Kolb and M.~S. Turner,
  \href{http://dx.doi.org/10.1201/9780429492860}{{\em {The Early Universe}}},
  vol.~69.
\newblock Taylor and Francis, 5, 1990.

\bibitem{ParticleDataGroup:2024cfk}
{\bfseries Particle Data Group} Collaboration, S.~Navas {\em et~al.}, ``{Review
  of particle physics},''
  \href{http://dx.doi.org/10.1103/PhysRevD.110.030001}{{\em Phys. Rev. D}
  {\bfseries 110} (2024) 030001}.

\bibitem{Baumann:2022mni}
D.~Baumann, \href{http://dx.doi.org/10.1017/9781108937092}{{\em {Cosmology}}}.
\newblock Cambridge University Press, 7, 2022.

\bibitem{Blasi:2022ayo}
S.~Blasi, A.~Mariotti, A.~Rase, A.~Sevrin, and K.~Turbang, ``{Friction on ALP
  domain walls and gravitational waves},''
  \href{http://dx.doi.org/10.1088/1475-7516/2023/04/008}{{\em JCAP} {\bfseries
  04} (2023) 008}, \href{http://arxiv.org/abs/2210.14246}{{\ttfamily
  arXiv:2210.14246 [hep-ph]}}.

\bibitem{Blasi:2023sej}
S.~Blasi, A.~Mariotti, A.~Rase, and A.~Sevrin, ``{Axionic domain walls at
  Pulsar Timing Arrays: QCD bias and particle friction},''
  \href{http://dx.doi.org/10.1007/JHEP11(2023)169}{{\em JHEP} {\bfseries 11}
  (2023) 169}, \href{http://arxiv.org/abs/2306.17830}{{\ttfamily
  arXiv:2306.17830 [hep-ph]}}.

\bibitem{Kawano:1989mw}
L.~Kawano, ``{Evolution of Domain Walls in the Early Universe},''
  \href{http://dx.doi.org/10.1103/PhysRevD.41.1013}{{\em Phys. Rev. D}
  {\bfseries 41} (1990) 1013}.

\bibitem{Avelino:2005kn}
P.~P. Avelino, C.~J. A.~P. Martins, and J.~C. R.~E. Oliveira, ``{One-scale
  model for domain wall network evolution},''
  \href{http://dx.doi.org/10.1103/PhysRevD.72.083506}{{\em Phys. Rev. D}
  {\bfseries 72} (2005) 083506},
  \href{http://arxiv.org/abs/hep-ph/0507272}{{\ttfamily arXiv:hep-ph/0507272}}.

\bibitem{Hiramatsu:2012sc}
T.~Hiramatsu, M.~Kawasaki, K.~Saikawa, and T.~Sekiguchi, ``{Axion cosmology
  with long-lived domain walls},''
  \href{http://dx.doi.org/10.1088/1475-7516/2013/01/001}{{\em JCAP} {\bfseries
  01} (2013) 001}, \href{http://arxiv.org/abs/1207.3166}{{\ttfamily
  arXiv:1207.3166 [hep-ph]}}.

\bibitem{Maggiore:2007ulw}
M.~Maggiore,
  \href{http://dx.doi.org/10.1093/acprof:oso/9780198570745.001.0001}{{\em
  {Gravitational Waves. Vol. 1: Theory and Experiments}}}.
\newblock Oxford University Press, 2007.

\bibitem{Kitajima:2023cek}
N.~Kitajima, J.~Lee, K.~Murai, F.~Takahashi, and W.~Yin, ``{Gravitational waves
  from domain wall collapse, and application to nanohertz signals with
  QCD-coupled axions},''
  \href{http://dx.doi.org/10.1016/j.physletb.2024.138586}{{\em Phys. Lett. B}
  {\bfseries 851} (2024) 138586},
  \href{http://arxiv.org/abs/2306.17146}{{\ttfamily arXiv:2306.17146
  [hep-ph]}}.

\bibitem{Kitajima:2023kzu}
N.~Kitajima, J.~Lee, F.~Takahashi, and W.~Yin, ``{Stability of domain walls
  with inflationary fluctuations under potential bias, and gravitational wave
  signatures},'' \href{http://arxiv.org/abs/2311.14590}{{\ttfamily
  arXiv:2311.14590 [hep-ph]}}.

\bibitem{Ferreira:2022zzo}
R.~Z. Ferreira, A.~Notari, O.~Pujolas, and F.~Rompineve, ``{Gravitational waves
  from domain walls in Pulsar Timing Array datasets},''
  \href{http://dx.doi.org/10.1088/1475-7516/2023/02/001}{{\em JCAP} {\bfseries
  02} (2023) 001}, \href{http://arxiv.org/abs/2204.04228}{{\ttfamily
  arXiv:2204.04228 [astro-ph.CO]}}.

\bibitem{Caprini:2019egz}
C.~Caprini {\em et~al.}, ``{Detecting gravitational waves from cosmological
  phase transitions with LISA: an update},''
  \href{http://dx.doi.org/10.1088/1475-7516/2020/03/024}{{\em JCAP} {\bfseries
  03} (2020) 024}, \href{http://arxiv.org/abs/1910.13125}{{\ttfamily
  arXiv:1910.13125 [astro-ph.CO]}}.

\bibitem{Caprini:2009fx}
C.~Caprini, R.~Durrer, T.~Konstandin, and G.~Servant, ``{General Properties of
  the Gravitational Wave Spectrum from Phase Transitions},''
  \href{http://dx.doi.org/10.1103/PhysRevD.79.083519}{{\em Phys. Rev. D}
  {\bfseries 79} (2009) 083519},
  \href{http://arxiv.org/abs/0901.1661}{{\ttfamily arXiv:0901.1661
  [astro-ph.CO]}}.

\bibitem{Cai:2019cdl}
R.-G. Cai, S.~Pi, and M.~Sasaki, ``{Universal infrared scaling of gravitational
  wave background spectra},''
  \href{http://dx.doi.org/10.1103/PhysRevD.102.083528}{{\em Phys. Rev. D}
  {\bfseries 102} (2020) 083528},
  \href{http://arxiv.org/abs/1909.13728}{{\ttfamily arXiv:1909.13728
  [astro-ph.CO]}}.

\bibitem{Ferreira:2023jbu}
R.~Z. Ferreira, S.~Gasparotto, T.~Hiramatsu, I.~Obata, and O.~Pujolas,
  ``{Axionic defects in the CMB: birefringence and gravitational waves},''
  \href{http://dx.doi.org/10.1088/1475-7516/2024/05/066}{{\em JCAP} {\bfseries
  05} (2024) 066}, \href{http://arxiv.org/abs/2312.14104}{{\ttfamily
  arXiv:2312.14104 [hep-ph]}}.

\bibitem{Dankovsky:2024zvs}
I.~Dankovsky, E.~Babichev, D.~Gorbunov, S.~Ramazanov, and A.~Vikman,
  ``{Revisiting evolution of domain walls and their gravitational radiation
  with CosmoLattice},''
  \href{http://dx.doi.org/10.1088/1475-7516/2024/09/047}{{\em JCAP} {\bfseries
  09} (2024) 047}, \href{http://arxiv.org/abs/2406.17053}{{\ttfamily
  arXiv:2406.17053 [astro-ph.CO]}}.

\bibitem{Manohar:2020nzp}
A.~V. Manohar and E.~Nardoni, ``{Renormalization Group Improvement of the
  Effective Potential: an EFT Approach},''
  \href{http://dx.doi.org/10.1007/JHEP04(2021)093}{{\em JHEP} {\bfseries 04}
  (2021) 093}, \href{http://arxiv.org/abs/2010.15806}{{\ttfamily
  arXiv:2010.15806 [hep-ph]}}.

\end{thebibliography}\endgroup
\end{document}